%% file: Thesis.tex
\documentclass[12pt]{ociamthesis}  

\usepackage{amssymb}

\title{Conformally Invariant Equations for Graviton     
        }   

\author{Mohsen Fathi}             
\college{Department of Physics,\\[1ex] Tehran Central Branch}  

\degree{Master of Science in Physics}     
\degreedate{Tehran, Winter 2010}         

\begin{document}

\baselineskip=18pt plus1pt

\setcounter{secnumdepth}{3} \setcounter{tocdepth}{3}

\maketitle                  
\include{dedication}        
\include{acknowledgements}   
\include{abstract}          

\begin{abstract}
Recent astrophysical data indicate that our universe might
currently be in a de Sitter (dS) phase. The importance of dS space
has been primarily ignited by the study of the inflationary model
of the universe and the quantum gravity. As we know Einstein's
theory of gravitation (with a nonzero cosmological constant) can
be interpreted as a theory of a metric field; that is, a symmetric
tensor field of rank-$2$ on a fixed de Sitter background. It has
been shown the massless spin-$2$ Fierz-Pauli wave equation (or the
linearized Einstein equation) is not conformally invariant. This
result is in contrary with what we used to expect for massless
theories. In this thesis we obtain conformally invariant wave
equation for the massless spin-$2$ in the dS space. This study is
motivated by the belief that conformal invariance may be the key
to a future theory of quantum gravity.
\end{abstract}

\begin{romanpages}          
\tableofcontents            
 \listoffigures             
\end{romanpages}            

\include{chapter1}
\include{chapter2}

\addcontentsline{toc}{chapter}{Introduction}
\include{Introduction}
\chapter{The Lorentz and the conformal groups, and the concept of
invariance}
\section{Group theory}
In order to explain symmetries in physics, it is common to use
groups, therefore the important properties of the groups, must be
defined.
\newtheorem{definition}{Definition}

\begin{definition}
A set of elements $\{a,b,c,...\}$ form a group, if there exist a
linear combination $a\circ b$ of these elements (called the
product), such that the following properties are confirmed for the
set:
\begin{itemize}
\item{the group $G=\{a,b,c,...\}$ comprises the product $ab$.}
\item{the set $G=\{a,b,c,...\}$ comprises a unit element $e$ such
that $ea=ae$.} \item{for every element of the group, there exist
an inverse element, such that $a^{-1}=a a^{-1}=e$.} \item{the
contribution property:$$(ba)c=a(bc).$$}
\end{itemize}
\label{def-1}
\end{definition}
Some peculiar virtues of groups have been listed below:
\begin{itemize}
\item{A group having the property $ab=ba$ for all of its elements,
is an Abelian group.}

\item{In a continuous group, the elements are functions of one or
several continuous variables, e.g. $G=\{a(t),b(t),...\}$, in which
$t$ is a continuous variable. }

\item{In any sequence, inside a compact group, there exists an
infinite number of partial sequences, converging to an element of
the group:
$$\lim_{n-\infty}a_n=a,\,\,\,\,\,\,\,\,\,\,\,\,\,\,\,\,a\in G.$$ }

\item{Two groups $\{a,b,c,...\}$ and $\{a',b',c',...\}$, are
isometric, if there exists a bijective transformation between
their elements (say $a\leftrightarrow a'$ and $b\leftrightarrow
b'$) such that \cite{1}:$$ab\leftrightarrow a'b'.$$}
\end{itemize}

Now we introduce two important groups.

\subsection{Orthogonal groups}
In mathematics, an orthogonal group of degree $n$ on field $F$
(notated by $O(n,F)$), is a group of $n\times n$ matrices. The
elements of these matrices come from $F$, and the group action is
done via matrix multiplication. This group is a subgroup of the
general linear group $GL(n,F)$, which is defined by:
\begin{equation}
O(n,F)=\{Q\in GL(n,F)|Q'Q=QQ'\}. \label{1-1}
\end{equation}
The classic orthogonal group on real numbers, is notated by
$O(n)$. The determinant of any orthogonal matrix is $\pm 1$. The
matrices with determinant $+1$, form a normal subgroup of the
orthogonal group, named the special orthogonal group $SO(n,F)$.
Both of these groups are algebraic groups. When the field $F$, is
a set of real numbers, these groups are shown simply by $O(n)$ and
$SO(n)$. Also these groups, form the compact Lie groups of
dimension $\frac{n(n-1)}{2}$.

These geometric properties are comprised by orthogonal and the
special orthogonal real groups:
\begin{itemize}
\item{The orthogonal group is a subgroup of the Euclidian group
$E(n)$, formed by isometries from $\mathbb{R}^n$, keeping the
origin unchanged. This group is the symmetric group of the
hyper-sphere $s^3$, and all the spherically symmetric geometrical
objects, if the origin is kept unchanged.}

\item{$SO(2,\mathbb{R})$ is a subgroup of $E(n)$, consist of
direct isometries, i.e. the isometries, keeping the origin and
directions unchanged. This group is indeed the rotation group for
all spherically symmetric geometrical objects, if the origin is
considered at their center.}

\item{$\{+I,-I\}$ is a normal subgroup of $O(n,\mathbb{R})$, and
if $n$ is an even number, this group also is a subgroup of
$SO(n,\mathbb{R})$. If $n$ is taken to an odd number,
$O(n,\mathbb{R})$ is the direct product of $SO(n,\mathbb{R})$ and
$\{+I,-I\}$. The rotation group $C_k$, formed by $k$ number of
rotations ($k$ is a correct number), is a normal subgroup of
$O(2,\mathbb{R})$ and $SO(2,\mathbb{R})$. }
\end{itemize}

Proportional to proper orthogonal basis, the isometries have the
following matrix form:

$$\left| {\begin{array}{*{20}{c}}
  {{R_1}}&{}&{}&{}&{}&{} \\
  {}& \ddots &{}&{}&0&{} \\
  {}&{}&{{R_k}}&{}&{}&{} \\
  {}&{}&{}&{ \pm 1}&{}&{} \\
  {}&0&{}&{}& \ddots &{} \\
  {}&{}&{}&{}&{}&{ \pm 1}
\end{array}} \right|.$$
The circular symmetric group is $O(2,\mathbb{R})$, from which
$SO(2,\mathbb{R})$ as a Lie group, is isomorphic to the sphere
$s^1$. This isomorphism, takes the complex number $\exp
(i\phi)=\cos (i\phi)+i\sin(i\phi)$ to the orthogonal matrix
\cite{2}:
$$\left|
{\begin{array}{*{20}{c}}
{\cos(\phi)}&{-\sin{(\phi})}\\
{\sin{(\phi})}&{\cos(\phi)}
\end{array}}
\right |.
$$

\subsection{Rotation groups}
In classical mechanics, the isotropy of space, means the
invariance of nature's rules. Let us consider two vectors
$\overrightarrow{r}$ and its rotated version $\overrightarrow{r'}$
in three dimensional space. We can write:
$$\overrightarrow{r'}=R\overrightarrow{r}.$$
One can state that the three basis of space, namely $e_1,e_2,e_3$,
rotate, and are mapped to $e'_1,e'_2,e'_3$. i.e.
\begin{equation}e'_i=R_{ij}e'_j. \label{2-1}\end{equation}
The $3\times 3$ rotation matrix, must be real. The inverse
transformation can be written as:
\begin{equation}e_i=U_{ij}e'_j. \label{3-1}\end{equation}
Therefore we will have:
\begin{equation}
\overrightarrow{r}=x_ie_i, $$$$
\overrightarrow{r'}=x_ie'_i=x'_ie_i, \label{4-1}
\end{equation}
or
\begin{equation}
x_ie'_i=x'_i(R^{-1})_{ij}e'_j, \label{5-1}
\end{equation}
in which
\begin{equation}
x_i=R_{ij}x_j. \label{6-1}
\end{equation}
The rotation matrices exhibit orthogonal properties:
\begin{equation}
R_{ij}R'_{ij}=\delta_{ii'},$$$$ R_{ij} R'_{ij}=\delta_{jj'},
\label{7-1}
\end{equation}
\begin{equation}
\textmd{det}(R_{ij}R'_{ij})=\textmd{det}(R_{ij})\textmd{det}(R'_{ij})=\textmd{det}(\delta_{jj'})$$$$\,\,\,\,\,\,\,\,
\Rightarrow\,\,\,\,\,\,\,\,\,
\textmd{det}^2(R_{ij})=1\,\,\,\,\,\,\,\,
\Rightarrow\,\,\,\,\,\,\,\,\, \textmd{det}(R_{ij})=\pm1.
\label{8-1}
\end{equation}
Therefore, the rotation will be separated into two distinct sets.
One set is a group which is formed by matrices of determinant
$+1$, and another is the matrices of determinant $-1$, which do
not form a group. Since the rotations are described by three
independent parameters, the so-called set, constitutes a
continuous three parameter group. This rotation group is notated
by $SO(3)$, showing that there exists an orthogonal group in three
dimensions, comprising all $3\times 3$ matrices of determinant
$+1$
\cite{3}.\\

The three dimensional isometries, keeping the origin unchanged,
are listed below:
\begin{itemize}
\item{The identity $I$.}

\item{Rotation around an axis, which is crossing the origin.}

\item{Rotation around an axis with an angle other that
$180^\circ$, which is combined with the reflections from the
origin surface.}

\item{An inverse in the origin.}

\item{A reflection from an origin crossing surface.}

\end{itemize}
The fourth and the fifth items, in a special case, and also the
sixth item, are known as special rotations \cite{4}.\\

As it was mentioned from mathematical view points, a symmetry is
represented by a group. If the symmetric properties of a system is
represented by a group $G$, a function (or some functions)
describes the system, possessing a part to represent the symmetry,
and another part to represent a real system. For example, if a
system $A$ of differential equations, accepts a symmetry group, it
would be so rare to reduce this system to a smaller system $B$
with a definite symmetry. Indeed, the system $B$, is derived from
$A$, by eliminating the symmetry \cite{8}.

\section{Invariance under a group action}
\begin{definition}
Let $G$ be a group of transformations on manifold $\mathcal{M}$.
The set $S\subset\mathcal{M}$ is called $G$-invariant, and $G$ is
called a symmetry group of S, if for a defined $g.p$ ($p\in S$ and
$g\in\mathcal{M}$) we have $g.p\in S$. \label{def-2}
\end{definition}

\begin{definition}
Let $G$ be a group of transformations on manifold $\mathcal{M}$.
The map
$$F:\mathcal{M}\longrightarrow \mathcal{N}$$
in which $\mathcal{N}$ is another manifold, is called a
$G$-invariant map, if for all $p\in \mathcal{M}$ and
$g\in\mathcal{N}$ which $g.p$ has been defined, $F(g.p)=F(p)$. A
$G$-invariant real valued function, is simply called invariant.
\label{def-3}
\end{definition}
\newtheorem{theorem}{Theorem}
\begin{theorem}
Let $G$ be a group of transformations on manifold $\mathcal{M}$. A
real valued function $f:\mathcal{M}\longrightarrow \mathbb{R}$ is
$G$-invariant, if and only if for all $p\in\mathcal{M}$ and
generators $\xi\in\mathbb{R}$ we have \cite{9}:
$$\xi_\mathcal{M}|_p(f)=0.$$
 \label{theo-1}
\end{theorem}

\subsection{Invariance of the laws of physics}
In 1872, Felix Klein proposed \emph{Erlangen Program f\"{u}r
Geomtry}, based on the group of symmetric transformations. In
1952, Fantappi\`{e}, having the same idea and using expressions
with a relativistic tendency, proposed \emph{Erlangen Program
f\"{u}r Physik}, representing a distinct worlds by a symmetric
group, keeping the laws of physics, invariant. It must be note
that this world, is indeed a physical system which is defined by a
symmetric group.

Isotropy of spacetime and its homogeneity with respect to laws of
physics, assert that these laws are based on symmetry. Therefore
it is believed that the laws of physics can be separated using
some symmetric groups, which leave them invariant. To do this, one
can use two groups, describing two different physical worlds. The
Galilei group and the Lorentz-Poincar\`{e} group.

Note that the Galilei group is an special case of Lorentz group in
$c\rightarrow\infty$. Mathematically the Lorentz group can be
expressed as a rotation-translation group, in a way that a
geometrical object, say Minkowski spacetime, is leaved invariant.

\section{The Lorentz group}
The three dimensional rotations in classical and quantum
mechanics, can be investigated using the group of transformations,
which keep the measurements constant. In special theory of
relativity, the Lorentz transformations of the 4-dimensional
coordinates $(x_0,\overrightarrow{x})$, obeys the following
invariance:
\begin{equation}
s^2=x_1^2+x_2^2+x_3^2-x_4^2. \label{9-1}
\end{equation}
Hence, the cinematics of special relativity, can be reexpressed by
the group of transformations, which keep $s^2$ invariant. This
group is the homogenous Lorentz group, consisting of normal
rotations, as well as the Lorentz group in Minkowski spacetime.
The group of transformations which create the invariance
$$s^2(x,y)=(x_0-y_0)^2-(x_1-y_1)^2-(x_2-y_2)^2-(x_3-y_3)^2,$$
are called the inhomogenous Lorentz group or the Poincar\`{e}
group. This group consists of translations and reflections in
spacetime.

The Lorentz group is a subgroup of the Poincar\`{e} group,
composed of all the isometries, which keep the origin unchanged.
Mathematically, the Lorentz group is the generalized orthogonal
group $O(1,3)$. This group is a 6-dimensional un-compact Lie
group, and one of its subgroups is $O(3)$ \cite{10,11}.

\section{The conformal group}
The isometric orthogonal transformations (preserving the
distances), also keep the angles, and consequently are conformal
maps. Nevertheless all the linear conformal transformations, are
not orthogonal. In the next chapter, these maps will be discussed
explicitly.

The group of conformal mappings in $\mathbb{R}^n$, is denoted by
$CO(n)$, consisting of the product of the orthogonal group, by the
dilation group. If $n$ is an odd number, these two groups do not
intersect, and indeed are the following product:
\begin{equation}
CO(2n+1)=O(2n+1)\times\mathbb{R}. \label{10-1}
\end{equation}
Whereas if $n$ is an even number, these two groups will intersect
and therefore (\ref{10-1}) is not a direct product. Instead it
will be a product by the subgroup of the dilation group with a
positive scalar, i.e.
\begin{equation}
CO(2n)=O(2n)\times\mathbb{R^+}. \label{11-1}
\end{equation}
Also one can define \cite{4}:
\begin{equation}
CSO(2n):=CO(n)\cap LG_+(n)=SO(n)\times\mathbb{R^+}. \label{12-1}
\end{equation}

\subsection{Conformal symmetry}
It is well-known that the conformal group is more powerful in two
dimensions than it in higher ones. In two dimensions, this group
possesses infinite dimensions. When the space has an Euclidian
metric, any function obeying the Cauchy-Riemann equations, will be
locally conformally invariant. If we consider the definition of
the conformal symmetries, same as the way that general relativity
keeps the light cone invariant, we find out that any
two-dimensional metric can be applied in this form. We can find a
transformation, to transform the so-called metric, to the same
metric, multiplied by a function. This property is not valid for
higher dimensions.

One of the definitions for the conformal symmetry, comes through
the group theory. This definition coincides the previous one, when
the space possesses more than two dimensions. To realize the
conformal symmetry, we ought to understand the mathematical field
operators, which are satisfying the Lagrange equations, which are
used to define the conformal group. These fields can be defined on
a Minkowskian manifold $x_\mu$, or on a six-dimensional imaginary
manifold $\eta_A$, correlated to $x_\mu$.\\

The conformal group of spacetime, is organized of coordinate
transformations, as listed below:
\begin{itemize}
\item{dilation
$$x_\mu=x\rho_\mu \,\,\,\,\textmd{where}\,\,\,\, \rho>0,$$}
\item{special conformal
transformations$$\sigma(x)=1-2cx+c^2x^2,\,\,\,\,\,\,$$$$x'_\mu=\sigma^{-1}(x)(x_\mu-C_\mu
x^2),$$} \item{inhomogenous Lorentz transformations.}
\end{itemize}
Here, we notate the conformal group by $SO(d,2)$, in which $d$ is
dimensions of spacetime. This transformation can be expressed as
the complete Poincar\`{e} group, which is a re-scaling
transformation. Also $\mu$ is a constant, and the special
conformal transformation would be:
\begin{equation}
x^\mu\longrightarrow\frac{x^\mu-b^\mu x^\mu}{1-2bx+xb^2x^2}.
\label{13-1}
\end{equation}
This group has $\frac{(d+2)(d+1)}{2}$ parameters and
\begin{equation}
x^2=x_\mu x^\mu=b_\mu x^\mu. \label{14-1}
\end{equation}
These special conformal transformations, are generated by
$R_\mu=P_\mu+IP_\mu I$, where $P_\mu$ is the translation
generator. The relation (\ref{13-1}), takes $x$ to
$\frac{-x}{x^2}$ \cite{5,6}.\\

The conformal group $SO(d,2)$ is generated by the Killing
equations; like the group of motions, for a five-dimensional
pseudo-sphere:
\begin{equation}
|\psi_1|^2+|\psi_2|^2-|\psi_3|^2=|\psi_4|^2=1$$$$\textmd{where}\,\,\,\,\psi=(\psi_1,\psi_2,\psi_3,\psi_4)\in
\mathbb{C}^4, \label{15-1}
\end{equation}
which has been solved accurately in spherical coordinates.
Therefore, the fifteen natural generators of the conformal group,
can be figured out, in the natural coordinates of this space. It
has been shown that the square operator $P_\mu P^\mu$, would not
be diagonal, when it is used to express the conformal group in the
Hilbert space of the harmonic functions on a five-dimensional
pseudo-sphere. Hence, to define the conformal group in the Hilbert
space of the harmonic functions on a projective five-dimensional
space, we write:
\begin{equation}
\psi^\dagger\beta\psi=0, $$$$ \textmd{and}\,\,\,\,\,\beta=\left|
{\begin{array}{*{20}{c}}
  {{1}}&{0}&{0}&{0} \\
  {0}& {1} &{0}&{0} \\
  {0}&{0}&{{-1}}&{0} \\
  {0}&{0}&{0}&{ -1}
\end{array}} \right|,
\label{16-1}
\end{equation}
contained in the complex space $\mathbb{C}^4$, and $\psi$ is a
vector. With
$$\psi=(\psi_1,\psi_2,\psi_3,\psi_4)$$
and
$$|\psi_4|^2=1,$$
and the substitutions:
$$\psi_1=x^1+xi^2,\,\,\,\,\,\psi_2=x^3+xi^6,\,\,\,\,\,\psi_3=x^4+xi^5,$$
we find out that $SO$ can be written as follows:
\begin{equation}
g_{\alpha\beta}x^\alpha x^\beta-(x^5)^2+(x^6)^2=1, \label{17-1}
\end{equation}
in which $g_{\alpha\beta}$ is the Minkowski metric, conventionally
having the signs $(+ + + -)$. Therefore $SO$ exhibits the
following fundamental form, in a six-dimensional space:
\begin{equation}
ds^2=g_{\alpha\beta}x^\alpha x^\beta-(x^5)^2+(x^6)^2. \label{18-1}
\end{equation}
It is known that the conformal group can be described as the group
of motions, in a five-dimensional pseudo-spherical space. The most
natural generators of this group, can be derived by solving the
Killing equations, for one dimensional motions in spherical
coordinates. Also the generators of Poincar\`{e} group in this
coordinates, are created by inserting the Minkowsi space in $SO$
\cite{7}.

\section{Invariance under Lorentz group (Lorentz invariance)}
The Lorentz invariance, is an expression in physics, due to the
properties of spacetime, which implies that:\\

\emph{In two different frames, which are having relative motions
and observing a same event, the non-gravitational laws of physics,
must have the same predictions about that event.}\\

A physical quantity is called Lorentz covariant, if it is
transformed under a given Lorentz representation. Such quantities,
if left invariant under the so-called transformations, would be
Lorentz invariant.

Due to the representation theory of Lorentz group, the Lorentz
covariant quantities, consist of scalars, 4-vectors, 4-tensors and
spinors. The spacetime interval, is a Lorentz invariant quantity,
just like the Minkowski norm for any 4-vector.

The correct equations in any inertial frame, are also Lorentz
covariant, which can be written due to the Lorenz covariant
quantities. According to the principal of relativity, all
fundamental laws of physics, must be Lorentz covariant.

Just think for a moment that the theory of relativity, would have
been derived form the following rational statement:\\

\emph{The newtonian physics, is invariant under rotation
$L_i=\epsilon_{ijk}x_jp_k$ and the Galilean transformations
$\overrightarrow{g_i}=t\overrightarrow{p_i}$.}\\

The mentioned generators, obey the following relations:
\begin{equation}
[L_i,L_j]=\epsilon_{ijk}L_i,
$$$$[L_i,g_j]=\epsilon_{ijk}g_k,$$$$[g_i,g_j]=0.
\label{19-1}
\end{equation}
To investigate these relations, it is easier to add a term
proportional to $\frac{1}{c^2}\epsilon_{ijk}L_k$, to the left hand
side of the last equation. The coefficient $c$ has the velocity
dimensions. In fact, $\overrightarrow{L}$ is dimensionless and
$\overrightarrow{g}$ has the inverse velocity dimension.
Afterwards one can construct a theory, to be invariant with
respect to the new group of motions. The different of this group
from Newtonian mechanics, arises only in speeds closed to $c$
\cite{12,21}.

\section{Conformal invariance}
Conformal invariance, has been proposed correlated to the scale
invariance, since the last century. For example,  the vacuum
Maxwell equations are both scale and conformal invariant. This
concept in quantum field theory, is illustrated by a local
energy-momentum tensor.

However, the applications of conformal invariance, have been
recognized in 1970. Till then it seemed that the consequences of
conformal invariance in spaces with arbitrary dimensions, were not
interesting. This will be quite different, when a two-dimensional
space is considered, since for two dimensions, the conformal group
possesses infinite dimensions.

In a word, conformal invariance, is a logical expansion of scale
invariance. The scale invariance, is dependent to the invariance
of our system, under the changes in homogenous length scales.
However, conformal invariance allows inhomogenous and local
changes in scales and the only need here is the invariance of
angles. This expansion, is indeed coherent, since it can be shown
that an invariant system under translations and rotations, at
least at continuity limits, would be scale invariant and exposes
short range effects. Therefore, the conformal symmetry is
spontaneously generated.

For two-dimensional systems, conformal symmetry notably has
improved our knowledge about nature. Whereas the scale invariance
can only put our system in large parts, dependent to properties
like symmetry, space dimensions and number of order parameters,
the conformal symmetry proposes and assortment of partition
functions.

Currently, researches on theoretical physics, show a considerable
focus on conformal invariance. However to use the technical tools
to understand it, one should have professional skills; lots of
initial concepts in string theory, require adequate knowledge of
quantum theory of fields. On the other hand we believe that it is
time to apply some methods in physics, based on conformal
invariance. While conformal invariance constitutes the basis of
higher dimensional string theory, it provides the possibility to
study the fields on oscillating metrics. This studies are of great
importance in two-dimensional quantum gravity \cite{20}.

\chapter{Conformal transformations}
\section{Conformal mapping}
\begin{figure}[htp]
\center\includegraphics[width=8.5cm]{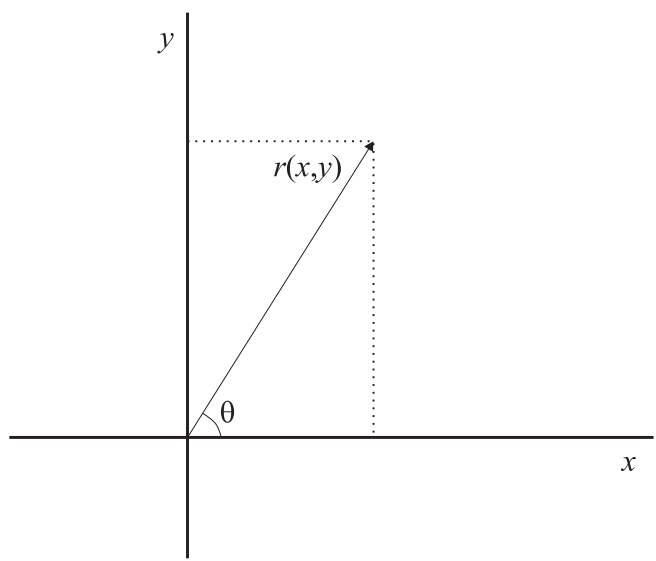} \caption{a complex number
in $x-y$ plane} \label{2.1}
\end{figure}

\begin{figure}[htp]
\center\includegraphics[width=14cm]{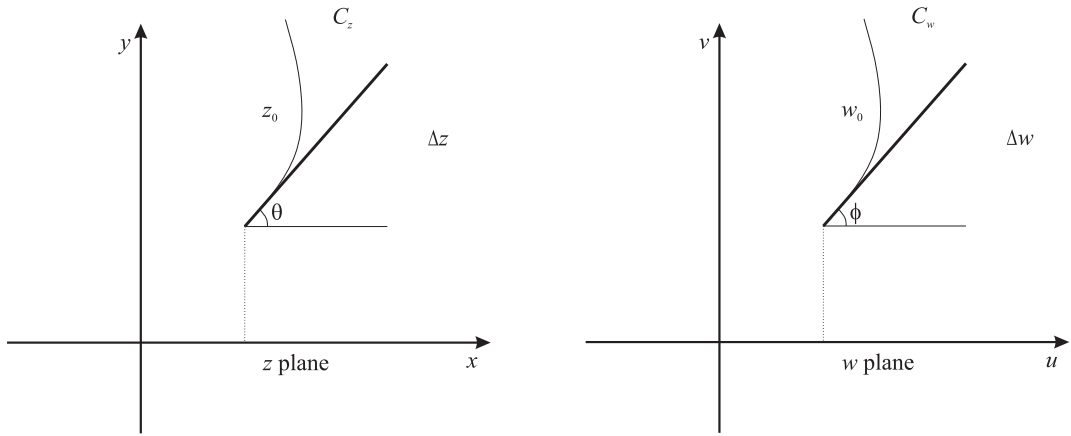} \caption{illustrating two
correspondent curves in complex planes} \label{2.2}
\end{figure}

Conformal mapping, or conformal transformation or the angle
preserving transformation, is indeed the transformation $w=f(z)$,
where $z$ would be the complex number $z=x+iy+$. Here $r=|z|$ is
the amplitude an $\theta=\textmd{arg}$ is the phase. $x$ and $y$
are the cartesian coordinates (see figure (\ref{2.1})).
\newtheorem{axiom}{Axiom}
\begin{axiom}
A function $f(z)$ is analytic at $z_0$, if it is differentiable at
$z=z_0$ and in its neighborhood. \label{ax-1}
\end{axiom}
As long as $f(z)$ is analytic, we have:
\begin{equation}
\frac{df}{dz}=\frac{dw}{dz}=\lim_{\Delta z\rightarrow
0}\frac{\Delta w}{\Delta z}. \label{1-2}
\end{equation}
Considering the polar version of this equation, it is possible to
equate the amplitudes and the phases with each other. For the
phases we can write:
\begin{equation}
\textmd{arg}\lim_{\Delta z\rightarrow0}\frac{\Delta w}{\Delta
z}=\lim_{\Delta z\rightarrow0}\textmd{arg}\frac{\Delta w}{\Delta
z}$$$$=\lim_{\Delta z\rightarrow0}\textmd{arg}\Delta
w-\lim_{\Delta z\rightarrow0}\textmd{arg}\Delta
z=\textmd{arg}\frac{dw}{dz}=\alpha,
 \label{2-2}
\end{equation}
in which $\alpha$ is the argument of the derivative, for a
constant definite $z$, and is independent of the direction from
which we have approached $z$. To declare the importance of
relation (\ref{2-2}), consider two curves, one $C_z$ in the $z$
plane, and an other curve $C_w$, corresponding to $C_z$ in the $w$
plane. As it is illustrated in figure (\ref{2.2}), the evolvement
$\Delta z$, makes an angle $\theta$ by the real axis $x$ and
corresponding to it, the evolvement $\Delta w$, makes an angle
$\phi$ by the real axis $u$.
From (\ref{2-2}) we conclude
\begin{equation}
\phi=\theta+\alpha.
 \label{3-2}
\end{equation}
In other words, as long as $w$ is analytic and its
differentiations are nonzero, any line in the $z$ plane, will be
rotated by an angle $\alpha$ in the $w$ plane. This conclusion is
valid for any $z_0$-crossing line. Therefore we can apply it for
two lines as well. The angle between these two lines would be:
\begin{equation}
\phi_2-\phi_1=(\theta_2+\alpha)-(\theta_1+\alpha)=\theta_2-\theta_1.
 \label{4-2}
\end{equation}
This relation asserts that the angle between two lines, will be
preserved by analytic transformations. Such transformations, which
keep the angles, are called conformal transformations. The
rotation angle $\alpha$, and also $|f(z')|$, are usually functions
of $z$.
\begin{axiom}
An analytic function, is conformal every where it possesses
nonzero differentiations. In reverse, any conformal mapping of a
complex variable, having continuous partial differentiations,
would be analytic. \label{ax-2}
\end{axiom}
As well as lots of other regions in physics, conformal mappings
are of great important in complex analysis.
\begin{definition}
A map which conserves the angles, but changes the directions, is
an isogonal map. \label{def-2}
\end{definition}
\begin{axiom}
A function $f:\mathbb{C}\longrightarrow\mathbb{C}$ is conformal,
if and only if there exist two complex numbers $a\neq0$ and $b$,
such that $$f(z)=az+b.$$ \label{ax-3}
\end{axiom}
Conformal transformations are rather profitable in solving physics
problems. Taking $w=f(z)$, the real and complex parts of $w(z)$,
must satisfy the Cauchy-Riemann and Laplace equations. Assume
$u(x,y)$ and $v(x,y)$ are respectively the real and the imaginary
parts of $f(z)$. Then:
$$\partial_xu=\partial_yv,\,\,\,\,\,\,\,\,\,\,\partial_vu=-\partial_xv\,\,\,\,\,\,\,\,\,\,\,\,\,\,\,\textmd{the Cauchy-Riemann condition},$$
$$\partial^2_xu+\partial^2_yu=0,\,\,\,\,\,\,\,\,\,\,\partial_x^2v+\partial_y^2v=0\,\,\,\,\,\,\,\,\,\,\,\,\,\,\textmd{the two-dimensional Laplace equation}.$$
Hence, spontaneously a scalar potential is provided. For a
three-dimensional electrostatic potential we have:
$$\partial^2_x\varphi+\partial^2_y\varphi+\partial^2_z\varphi=0,$$
which is reliable in a space without any charge density.
Generally, any function which satisfies the Laplace equation, is
called a \textit{harmonic} function. Therefore the above
electrostatic potential, is a three-dimensional harmonic function.

As stated above, the real and imaginary parts of a complex
function are harmonic, therefore complex analysis methods become
sometimes advantageous in solving electrostatic problems. Since
$\varphi$ satisfies the Laplace equation, one can consider it as a
part of the analytic function $w(z)$, which here we call it a
complex potential. To obtain the potential for a sequence of
charges, firstly we must derive $w(z)$ for a line of charges,
which embarks on moving when it is at the origin. If this line is
located at $z_0=x_0+iy_0$, then it can be shown that:
$$w(z)=2\lambda\ln(z-z_0),$$
where $\lambda$ is the linear charge density. And in general for a
sequence of charges we have:
\begin{equation}
w(z)=2\sum_{k=1}^n\lambda_k\ln(z-z_k).
 \label{5-2}
\end{equation}
The function $w(z)$ can be used to solve electrostatic problems,
with simple charge distributions.\\

Instead of discussing $w(z)$, let us consider a map from the $z$
plane (or $xy$ plane) to the $w$ plane (or $uv$ plane). In special
case, the curves with same potential, are mapped to lines parallel
to $z$ axis in the $w$ plane, since these curves are defined to be
$u$-constant. Also the $v$-constant curves are mapped to
horizontal lines in $w$ plane. This would be a big simplification
in geometry:\\

\emph{The straight lines, specially when they are parallel to the
axis, are far simpler than circles and are easier to be analyzed
than other geometrical objects. Specially when their centers do
not coincide to the origin.}\\

Therefore let us introduce two different complex worlds; one the
$x-y$ plane an the other, which is primed, would be the $x'-y'$
plane. Assume that we are in the $z$ coordinate
system\footnote{Here the $x-y$ plane is denoted by $z$, and $z'$
is used instead of $w$ and $(x',y')$ instead of $(u,v)$.} and we
wish to obtain a quantity like the electrostatic potential
$\varphi$. If solving this problem was hard in $z$, we could just
translate it to $z'$ to simplify the resolving process. To do
this, we solve the problem with respect to $(x',y')$ and then, we
translate it again to $z$. If any physical problem exists for
which this solution is valid, then the problem is solved.
Therefore pursuing an inverse procedure, we find a solution, for
which if we wanted to obtain straightforwardly, we had could have
confronted a sophisticated work. Hence, mappings that relate $z$
and $z'$, should be selected accurately. \\

We ought to consider two necessary conditions:
\begin{itemize}
\item{First Condition: The differential equation which is
describing the physics of our problem, must be simpler than a
translation in $z'$. Since the Laplace equation is among the
simplest ones, the $z'$ world should obey the Laplace equation.}

\item{Second Condition: The applied mapping must conserve the
angles. This condition is of great importance, since the
equipotential curves and the filed lines, are supposed to be
perpendicular to both $z$ and $z'$ worlds.}
\end{itemize}

Briefly speaking, our map must be conformal.

\begin{theorem}
Consider $\gamma_1$ and $\gamma_2$ be two curves in the complex
plane $z$, making an angle $\alpha$ at point $z_0$. Assume that
$f:\mathbb{C}\longrightarrow\mathbb{C}$ is a map with relation
$f(z)=z'=x'+iy'$, which is analytical at $z_0$. Also consider
$\gamma'_1$ and $\gamma'_2$ be transformed curves by $f$, making
the angle $\alpha'$ with each other. Hence, we will have:

\textmd{a}) $\alpha'=\alpha$, or the map is conformal
if$$\frac{df}{dz}|_{z=z_0}\neq0,$$

\textmd{b}) if $f$ is harmonic at $(x,y)$, therefore it will be
harmonic at $(x',y')$. \label{theo-2}
\end{theorem}

Some examples of conformal mapping are listed below:
\begin{itemize}
\item{$z'=z+a$, where $a$ is an arbitrary constant. This simply
can be regarded as a translation from the $z$ axis.}

\item{$z'=bz$}

\item{$z'=\frac{1}{z}$}

\item{combining the formerly mentioned transformations, we obtain
the following map:$$z'=\frac{az+b}{cz+d},$$ which will be
conformal if$$\frac{dz'}{dz}\neq0\neq cz+d.$$ Also when
$az+b\neq0$. }
\end{itemize}

Such mappings are often called \textit{homographic}
transformations. One of the advantages of these transformations is
that they can map an infinite domain from the $z$ plane, to a
finite domain in $z'$ plane. In fact using them we can transform
high valued points to a neighborhood of point $z'=\frac{a}{c}$.
\\\\\\
\textbf{Example: A physical application} \cite{Wolfram}\\

Consider the following potential:
$$w(z)=Az^n=Ar^n e^{in\theta}.$$
The real and the imaginary parts are respectively
$$\phi=Ar^n\cos(n\theta),$$
$$\psi=Ar^n\sin(n\theta).$$\\
\underline{For $n=-2$}:
\begin{figure}[htp]
\center\includegraphics[width=6cm]{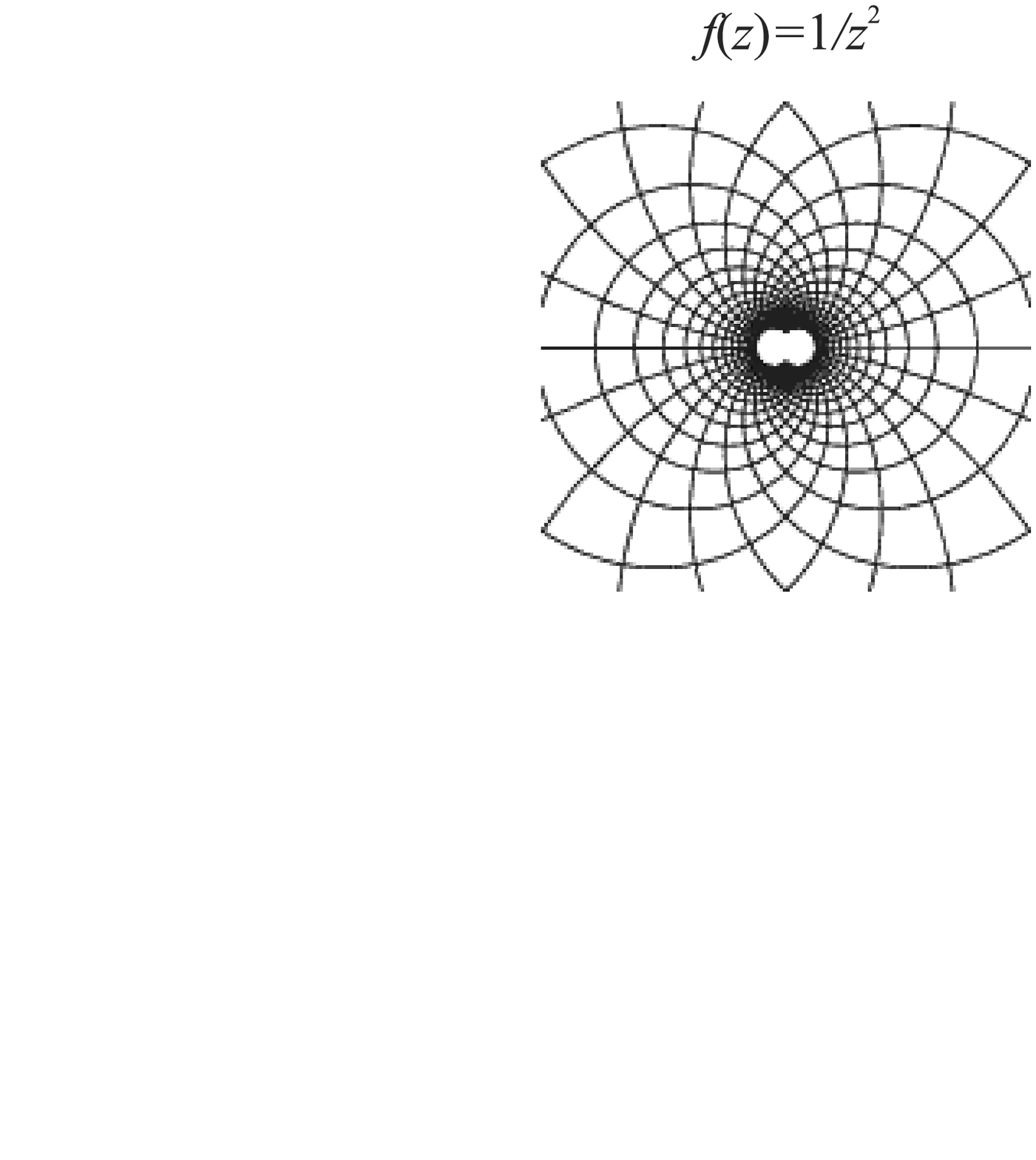} \caption{the equipotential
curves for $n=-2$}
 \label{3.2}
\end{figure}
The equipotential curves are shown in figure (\ref{3.2}) and the
two parts of the potential are derived as:
$$\phi=\frac{A}{r^2}\cos({2\theta}),$$
$$\psi=-\frac{A}{r^2}\sin(2\theta).$$
\\
\underline{For $n=-1$}:
\begin{figure}[htp]
\center\includegraphics[width=3.5cm]{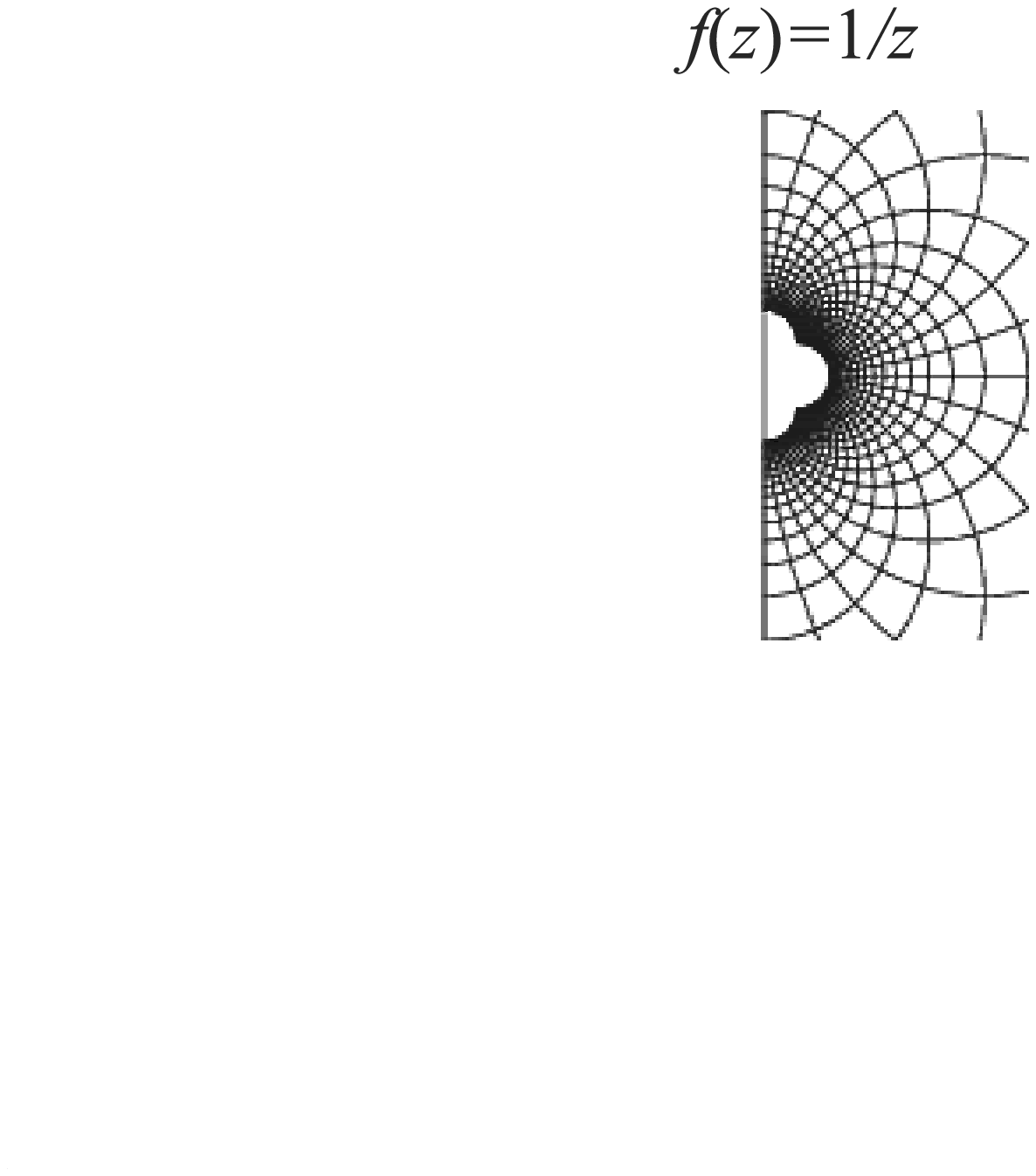} \caption{the equipotential
curves for $n=-1$}
 \label{4.2}
\end{figure}
The equipotential curves are shown in figure (\ref{4.2}) and the
two parts of the potential are derived as:
$$\phi=\frac{A}{r}\cos({\theta}),$$
$$\psi=-\frac{A}{r}\sin(\theta).$$
This solution is composed of two circular systems, and $\phi$ is
the potential function on two parallel lines of charge with
opposite signs.\\

\underline{For $n=\frac{1}{2}$}:
\begin{figure}[htp]
\center\includegraphics[width=4cm]{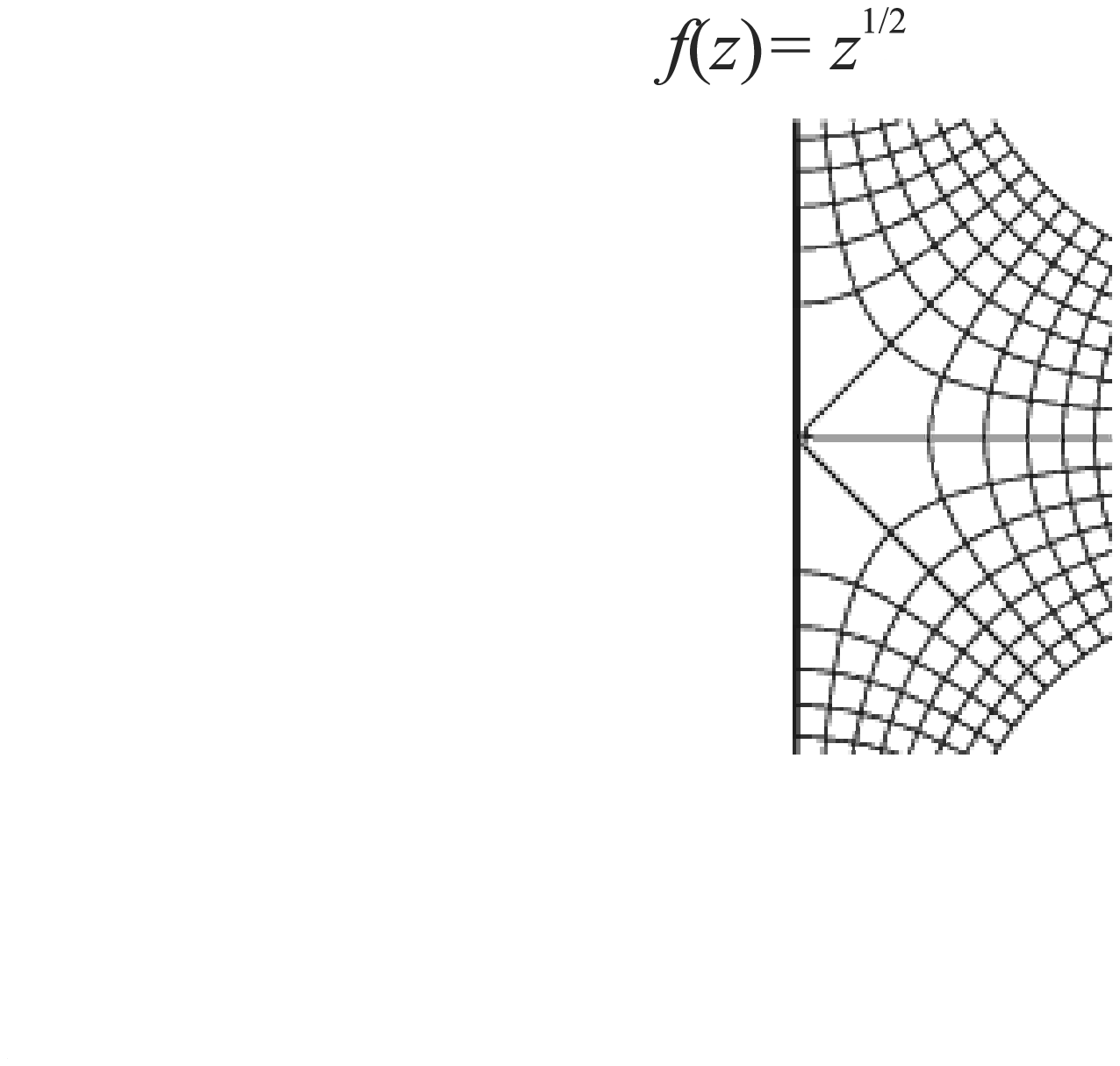} \caption{the equipotential
curves for $n=\frac{1}{2}$}
 \label{5.2}
\end{figure}
The equipotential curves are shown in figure (\ref{5.2}) and the
two parts of the potential are derived as:
$$\phi=Ar^{\frac{1}{2}}\cos(\frac{\theta}{2})=A\sqrt{\frac{(x^2+y^2)^{\frac{1}{2}}+x}{2}},$$
$$\psi=Ar^{\frac{1}{2}}\sin(\frac{\theta}{2})=A\sqrt{\frac{(x^2+y^2)^{\frac{1}{2}}-x}{2}},$$
from which, $\phi$ gives the field near the edge of a thin layer.
\\

\underline{For $n=1$}:
\begin{figure}[htp]
\center\includegraphics[width=5.5cm]{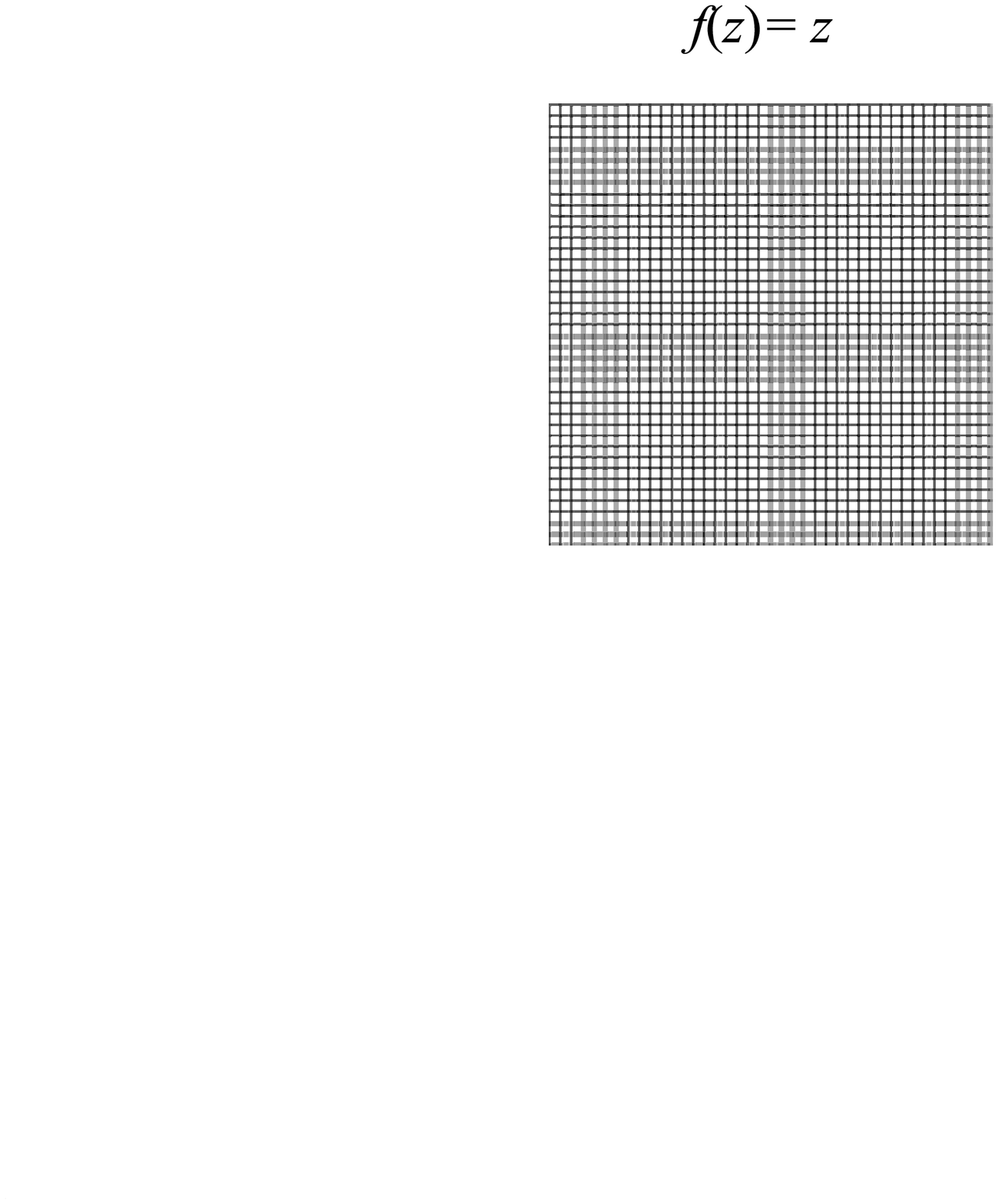} \caption{the equipotential
lines for $n=1$}
 \label{6.2}
\end{figure}
The equipotential curves are shown in figure (\ref{6.2}) and the
two parts of the potential are derived as:
$$\phi=Ar\cos(\theta),$$
$$\psi=Ar\sin(\theta),$$
which are two straight lines.
\\

\underline{For $n=\frac{3}{2}$}:
\begin{figure}[htp]
\center\includegraphics[width=5.5cm]{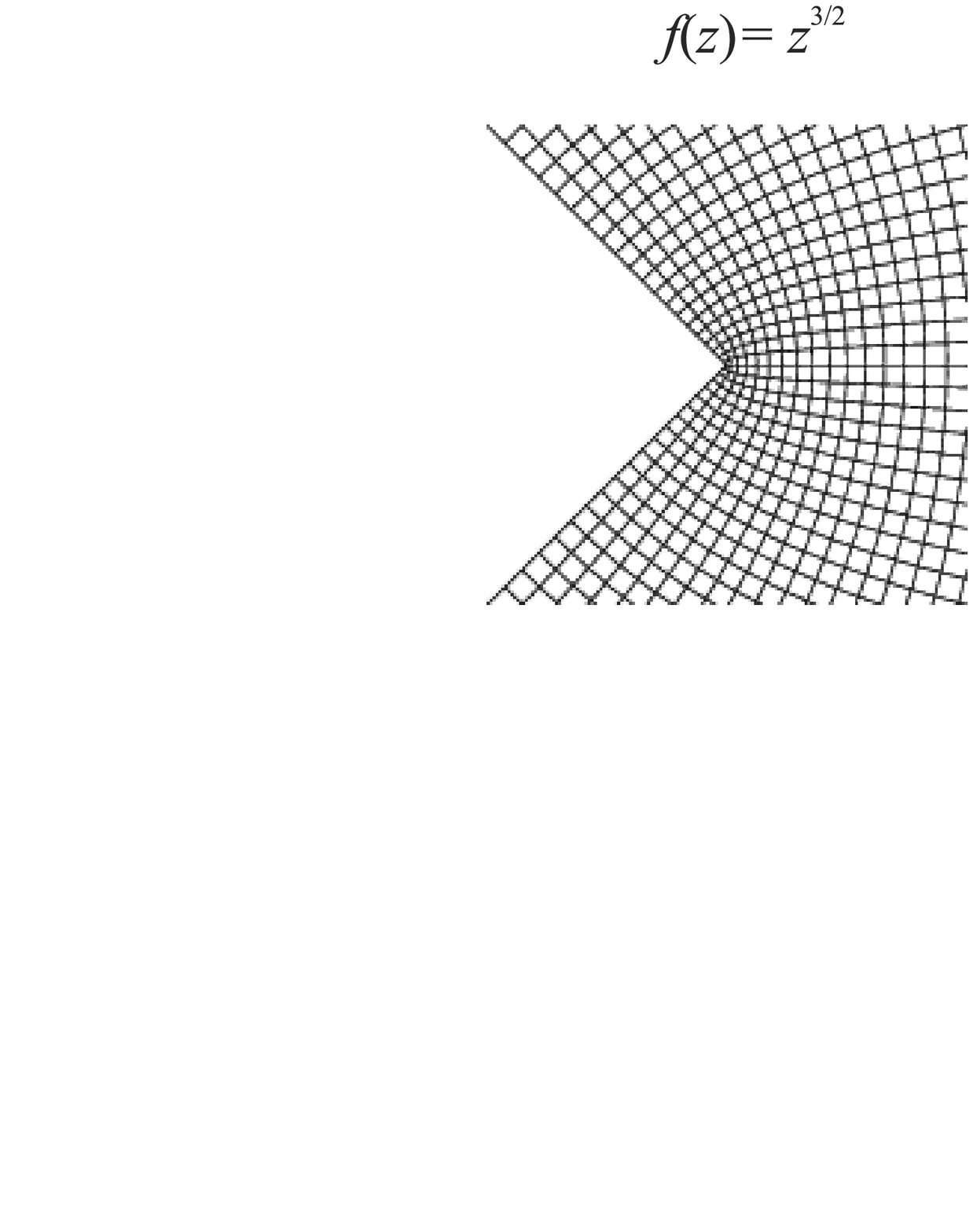} \caption{the equipotential
curves for $n=\frac{3}{2}$}
 \label{7.2}
\end{figure}
$\phi$ gives the field near and outside of a corner, made of
perpendicular plates. The equipotential curves are shown in figure
(\ref{7.2}). Also the potential is given by:
$$w=Ar^{\frac{3}{2}}e^{\frac{3}{2}i\theta}.$$
\\
\underline{For $n=2$}:
\begin{figure}[htp]
\center\includegraphics[width=5.5cm]{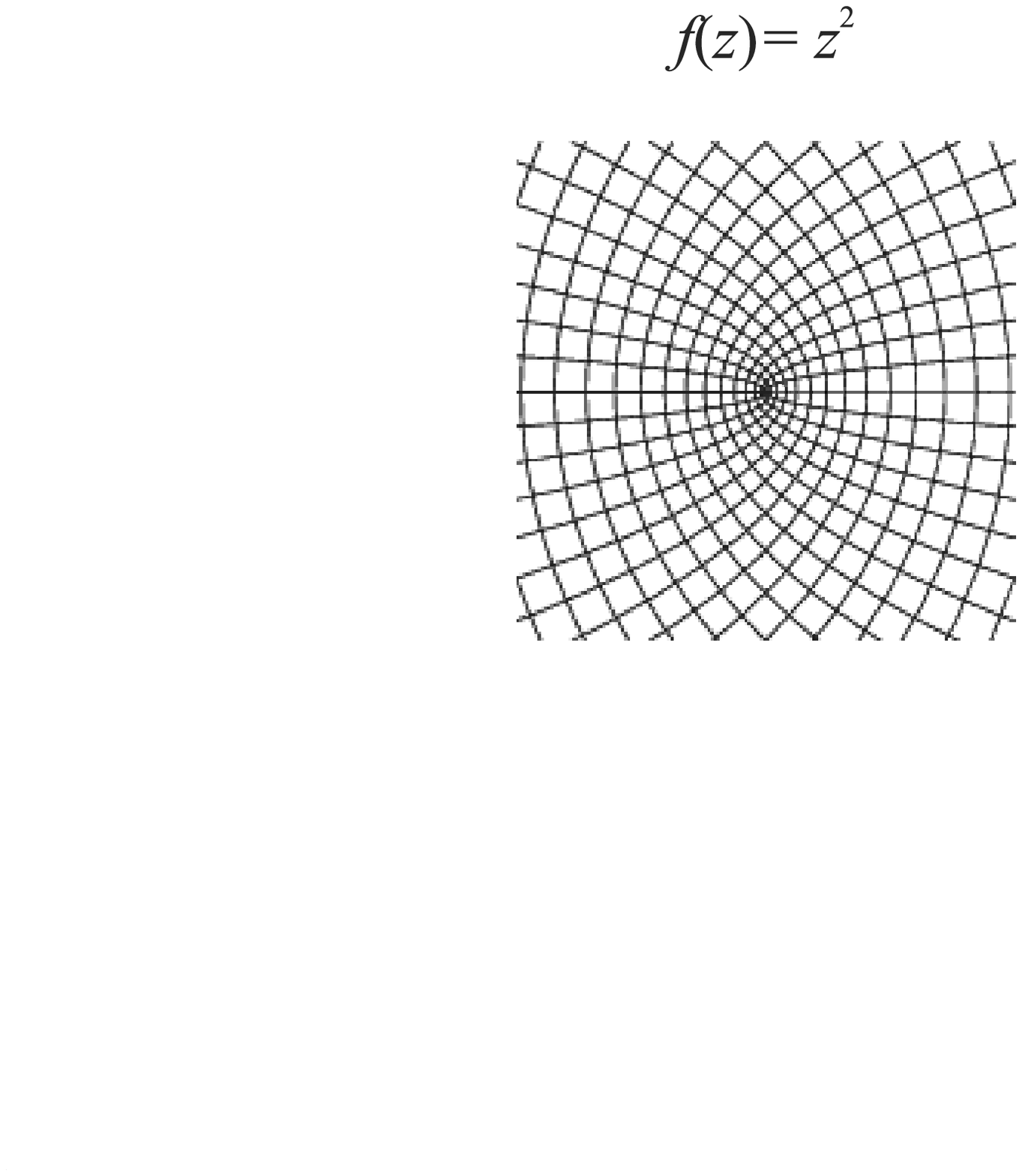} \caption{the equipotential
curves for $n=2$}
 \label{8.2}
\end{figure}
The potential and its real and imaginary parts are:
$$w=A(x+iy)^2=A[(x^2-y^2)+2ixy],$$
$$\phi=A(x^2-y^2)=Ar^2\cos(2\theta),$$
$$\psi=2Axy=Ar^2\sin(2\theta),$$
The equipotential curves are shown in figure (\ref{8.2}), which
are two perpendicular hyperboloids, and $\phi$ would be the
potential at the vicinity of the middle point, between two
separated charges. Also $\phi$ can be regarded as the potential on
the open part of a rectangular conductor \cite{9,16,22}.
\begin{flushright}
$\blacksquare$
\end{flushright}

\section{Conformal covariance and conformal invariance, and the
scalar wave equations} A new formulation of scalar field equations
is given by Klein-Gordon equation, when there exist a metric
tensor $g^{ij}(x)$ and a scalar field $C(x)$. This equation is
covariant with respect to the gauge transformations
\begin{equation}
A(x)\longrightarrow A_i(x)+\nabla_iV(x), \label{6-2}
\end{equation}
and the conformal transformations of the tensor field, i.e.
\begin{equation}
g^{ij}\longrightarrow \textmd{exp}[\theta(x)] g^{ij}, \label{7-2}
\end{equation}
in which $V(x)$ and $\theta(x)$ are arbitrary functions of
$x=x(x^1,x^2,...,x^n)$. In this case, the square of the remnant
mass $m^2(x)$ is defined is a function of given fields, in the
form:
\begin{equation}
m^2(x)=c-\frac{n-2}{4(n-1)}R-\frac{1}{4}A^iA_i-\frac{1}{2}\nabla_iA^i,
\label{8-2}
\end{equation}
which is transformed as follows:
\begin{equation}
m^2(x)\longrightarrow\textmd{exp}[-\theta(x)]m^2. \label{9-2}
\end{equation}
In equation (\ref{8-2}), $R$ is the scalar curvature, and
$\nabla_i$ is the covariant derivative in the Reimannian space
$V_n$. Also for the metric tensor $g_{ij}(x)$ (the inverse of
$g^{ij}(x)$) we have $A(x)\longrightarrow g_{ij}A^j$.

Considering a class of wave equations with constant mass, it has
been shown that we shall confront the transformation matrices
$\bar g^{ij}(x)\longrightarrow m^2(x)g_{ij}(x)$, depending
directly on the given tensor and scalar and vector fields. For the
vector field potential $\nabla_i A_j=\nabla_j A_i$, we have
equation (\ref{10-2}) which is the equation of motion of a free
particle in Riemannian space $V_n$.
\begin{equation}
g^{ij}\nabla_i\nabla_j\phi+\frac{n-2}{4(n-1)}R\phi+m_0^2\phi=0.
\label{10-2}
\end{equation}
This equation exposes the complete characteristics of a group, and
is comparable to the equation
\begin{equation}
g^{ij}\nabla_i\nabla_j\phi+m_0^2\phi=0, \label{11-2}
\end{equation}
which is considerable in quantum mechanical expansions and the
quantum theory of fields, where the space curvature is assumed to
be nonzero \cite{17}.

\section{Conformal transformations, and conformal invariance in
gravitation} The conformal transformations are used frequently in
studying the relations between diversified theories of
gravitation, and Einstein relativity. Therefore inevitably, it is
important to consider the instructions of conformal
transformations for geometrical quantities in general relativity.

In special case, we concern about the conformal transformations of
the energy-momentum tensor, and as a must, we should proceed with
the delicate and exact principal of conservation (or the Bianchi
identity) in one of the conformal systems, as an origin for the
others. The importance of this principal, goes back to this fact
that the conformal transformations create a matter term, composed
of the conformal factor, and insert it into the conservation
principal.

In the outstanding case of flat spacetime, this matter is created
from the work, done by the conformal transformations to curve the
spacetime.

We also have to note the structure of conformal gravity. In this
case which is the simplest one, we should concern about the
Brans-Dicke theory, taking its parameter to be $\frac{3}{2}$.

The conformal transformations are applied to investigate
gravitation in higher dimensional theories. In this way, we obtain
the laws of conformal transformations for scalar invariants,
namely $R^2$, $R_{ab}R^{ab}$ and $R_{abcd}R^{abcd}$, and
consequently for the Gauss-Bonnet invariant in arbitrary
dimensions.

Conformal transformations of a metric tensor, exhibits some
interesting characteristics of gravitational theories, which are
based on scalars. Here the point is that we can demonstrate these
theories, in two different systems with conformal relations:\\

1) The Jordan system, in which the scalar field is coupled by the
metric tensor, non-minimally.\\

2) The Einstein system, where this field has a minimum coupling
with the metric tensor.\\

Note that the scalar based gravitational theories , are the lower
limits of the super string theory. It has been shown that some
physical processes like expansion of the universe or the
perturbations in density, are seemed to be different in related
conformal systems. Therefore, it seems to be crucial to
investigate these transformations, through diversified
gravitational theories.

Lots os studies have been devoted to the problem of changes in
geometrical quantities under conformal transformations. However
the transformation laws, have not always been clarified and some
simple traits of them have not been investigated in details.
Therefore, we should discuss these laws explicitly, and also
concern about conformal transformations in higher-dimensional
curvatures.

\section{Conformal transformations in Einstein gravity}
Let us consider the spacetime $(\mu,g_{ab})$, where $\mu$ is a
smooth manifold and $g_{ab}$ is the Lorentz metric on $\mu$. The
conformal transformation
\begin{equation}
\bar g_{ab}(x)=\Omega^2 g_{ab}(x), \label{12-2}
\end{equation}
in which $\Omega$ is an uniform and nonzero function of spacetime.
The transformation in (\ref{12-2}) is a congruence of the metric,
independent of points, and is called the \textit{conformal
factor}. This factor relies in the interval $0<\Omega<\infty$.

The conformal transformations, increase or decrease the distances
between two definite points, in the same coordinate systems $x^a$
on manifold $\mu$, but leave the angles between the vectors,
unchanged. This leads to preservation of manifold's structure.
Assuming a constant $\Omega$, we will have a congruent
transformation. In fact, conformal transformations are local
congruences. We have:
$$\Omega=\Omega(x).$$
On the other hand, the coordinate transformations like
$x^a\longrightarrow \bar x^a$, only change the coordinate systems,
not the geometry. Consequently, such transformations differ from
conformal transformations. This is an accurate fact, since
conformal transformations, lead to new physical conditions.
Whereas this concept, relates to different couplings between
physical fields and gravitation, we investigate different systems,
in which the physics of our problem is studied.

In a $D$-dimensional spacetime, the determinant of metric, i.e.
$g=\textmd{det}(g_{ab})$ is transformed in follows:
\begin{equation}
\sqrt{-\bar g_{ab}}=\Omega^D\sqrt{- g_{ab}}. \label{13-2}
\end{equation}
From (\ref{13-2}) we have:
\begin{equation}
\bar g^{ab}=\Omega^{-2} g^{ab}, \label{14-2}
\end{equation}
and
\begin{equation}
\bar {ds^2}=\Omega^{2} ds^2. \label{15-2}
\end{equation}
Finally, the concept of conformal flatness will be:
\begin{equation}
\bar g_{ab}\Omega^{-2}(x)=\eta_{ab}, \label{16-2}
\end{equation}
where $\eta_{ab}$ is the Minkowski metric. Using conformal
transformations of the metric, we are able to calculate such
transformations of the Einstein tensor, after obtaining the
Christoffel symbols and Riemann and Ricci tensors, in
$D$-dimensional spacetime. We have:
\begin{equation}
\bar
G_{ab}=G_{ab}+\frac{D-2}{2\Omega^2}[4\Omega_{,a}\Omega_{,b}+(D+5)\Omega_{,c}\Omega^{,c}g_{ab}]-\frac{D-2}{\Omega}[\Omega_{,ab}-g_{ab}\Box\Omega].
\label{17-2}
\end{equation}
In this equation, the D'alembertian $\Box$ acts with respect to
$g_{ab}$. An important feature of conformal transformation is that
they preserve the Weyl curvature tensor
\begin{equation}
C_{abcd}=R_{abcd}+\frac{2}{D-2}[g_{a[d}R_{c]b}+g_{b[c}R_{d]a}]+\frac{2}{(D-2)(D-1)}Rg_{a[c}g_{d]b}.
\label{18-2}
\end{equation}
This means that from (\ref{12-2}) we conclude
\begin{equation}
\bar C_{abcd}=C_{abcd}. \label{19-2}
\end{equation}
Using (\ref{19-2}) together with relations (\ref{12-2}) and
(\ref{14-2}), one can simply derive the Weyl Lagrangian. This
Lagrangian is also invariant under conformal transformations, i.e.
\begin{equation}
\bar L_W=-\alpha(-\bar g)^{\frac{1}{2}}\bar C^{abcd}\bar
C_{abcd}=-\alpha(\bar g)^{\frac{1}{2}} C^{abcd}C_{abcd}=L_W.
\label{20-2}
\end{equation}

\section{Explaining some axioms using conformal transformations}
\begin{axiom}
Assume $\mathcal{M}$ be a complete Einstein manifold on which
there exists a vector field, generating a $1$-parameter group of
conformal transformations. Thereupon $\mathcal{M}$ is isometric to
a continuous space with positive curvature. In spacial case,
$\mathcal{M}$ is homeomorphic to the sphere $s^n$. \label{ax-4}
\end{axiom}
On the other hand, a pseudo-circular transformation from
Riemannian manifold $\mathcal{M}$ with metric $g_{\mu\lambda}$, to
the Riamannian manifold $\mathcal{M'}$ with metric
$g'_{\mu\lambda}$, is a conformal transformation, i.e.
\begin{equation}
g'_{\mu\lambda}=\rho^2 g_{\mu\lambda}. \label{21-2}
\end{equation}
This relation, takes the circular geodesics in $\mathcal{M}$ to
circular geodesics in $\mathcal{M'}$. This process can be denoted
by the following equation:
\begin{equation}
\nabla_\mu\rho_\lambda-\rho_\mu\rho_\lambda=\psi g_{\mu\lambda},
\label{22-2}
\end{equation}
where $\rho$ and $\psi$ are real-valued functions on
$\mathcal{M}$. Also
\begin{equation}
\rho_\lambda=\nabla_\lambda\log \rho. \label{23-2}
\end{equation}

\begin{axiom}
Assume $\mathcal{M}$ and $\mathcal{M'}$ be two Riemannian
manifolds with constant curvatures $k$ and $k'$. Consider
$\mathcal{M'}$ to be complete and also there exists a
pseudo-circular transformation from $\mathcal{M}$ to
$\mathcal{M'}$. Therefore $\mathcal{M}$
\begin{itemize}
\item{is the Euclidean space if $k=0$,}

\item{is a spherical space if $k>0$,}

\item{is a hyperbolic space if $k<0$.}

\end{itemize}
 \label{ax-5}
\end{axiom}

\begin{axiom}
In addition to the assumptions in axiom 5, assume that, is also
complete and the pseudo-circular transformation, is a homeomorphic
transformation from $\mathcal{{M}}$ to $\mathcal{M'}$. Therefore
the constant curvatures $k$ and $k'$ must be positive and
$\mathcal{{M}}$ and $\mathcal{M'}$ are spherical spaces.
 \label{ax-6}
\end{axiom}

We should note that:

\begin{theorem}
A conformal transformation, which transforms an Einstein manifold
to another one, is a pseudo-circular transformation.
\label{theo-3}
\end{theorem}

\begin{theorem}
If a complete Einstein manifold $\mathcal{M}$ is transformed
conformally to another Einstein manifold $\mathcal{M'}$, therefore
$\mathcal{M}$
\begin{itemize}
\item{is the Euclidean space if $k=0$,}

\item{is a spherical space if $k>0$,}

\item{is a hyperbolic space if $k<0$.}

\end{itemize}

\label{theo-4}
\end{theorem}

\begin{theorem}
If a complete Einstein manifold $\mathcal{M}$ accepts its
conformal transformation to itself, thereupon $\mathcal{M}$ would
be the spherical space \cite{19}. \label{theo-5}
\end{theorem}

\chapter{Einstein field equations}
\section{Introduction}
The Einstein field equations are a set of 10 equations in general
theory of relativity, which explain the gravitational reactions,
as a consequence of the spacetime curvature under the influence of
mass or energy \cite{24}.

Einstein equation was first proposed by Einstein in 1915, as a
tensorial equation. This equation equates the spacetime curvature
(which is described by Einstein tensor) to the energy (which is
described by energy-momentum tensor) \cite{25}.

Just like the procedure in which the electromagnetic fields are
determined by Maxwell's equations, using the charges and currents,
Einstein field equations are applied to determine the geometry of
spacetime, as a consequence of the presence of energy, mass and
linear momentum. This means that these equations, give the metric
for a definite form of energy in spacetime.

The correlations between the metric tensor and Einstein tensor,
demonstrates the field equations, as a set of partial differential
equations. The solutions to these equations, would be the metric
tensor components. Also the trajectories due to particle motions,
can be derived from the consequent geometry. Since Einstein field
equations are locally obeying energy-momentum conservation, these
equations reduce to Newtonian field equations, in weak field
limits.

\section{The mathematical form}
Einstein field equation can be written as follows \cite{24}:
\begin{equation}
R_{\mu\nu}-\frac{1}{2}g_{\mu\nu}R+g_{\mu\nu}\Lambda=\frac{8\pi
G}{c^4}T_{\mu\nu},\label{1-3}
\end{equation}
where $R_{\mu\nu}$ is the Ricci curvature tensor, $R$ the scalar
curvature, $g_{\mu\nu}$ the metric tensor, $\Lambda$ the
cosmological constant, $G$ the gravitational constant,
$T_{\mu\nu}$ the energy-momentum tensor and $c$ is the speed of
light.

Einstein field equation, is a tensorial equation which relates a
set of $4\times 4$ tensors. Each tensor possesses 10 independent
components. Considering our freedom in choosing the spacetime
coordinates, the number of independent equations decreases to 6.

Albeit Einstein equations had been initially based on
four-dimensional spacetime, some theorists have expanded the
results in $n$ dimensions. These equations, which are technically
regarded to be out of the region of general relativity, are still
referred to Einstein equations. The vacuum field equations, define
the concept of Einstein manifold.

Despite the simple form, Einstein equations are indeed rather
complicated. For an ordinary distribution of matter and energy,
Einstein field equations become a set of equations for the metric
tensor $g_{\mu\nu}$, on which the Ricci tensor and the Ricci
scalar, will non-linearly depend.

In fact, Einstein field equations, when fully written, are
combined of ten coupled non-linear hyperbolic-elliptical
differential equations.

Introducing the Einstein tensor
\begin{equation}
G_{\mu\nu}=R_{\mu\nu}-\frac{1}{2}g_{\mu\nu}R,\label{2-3}
\end{equation}
which is a symmetric tensor of rank two, one can write the
equations in the following compact form:
\begin{equation}
G_{\mu\nu}=\frac{8\pi G}{c^4}T_{\mu\nu},\label{3-3}
\end{equation}
where the cosmological term has been moved into the
energy-momentum tensor, pertaining the Dark energy concept. In
geometrical units $G=c=1$, this equation reduces to
\begin{equation}
G_{\mu\nu}=8\pi T_{\mu\nu},\label{4-3}
\end{equation}
in which the left hand side (lhs), indicates the spacetime
curvature, caused by the metric tensor. The right hand side (rhs)
stands for the matter/energy contents of spacetime. Einstein field
equations can be expressed as a set of equations, demonstrating
how the spacetime will curve, related to the matter/energy
constituting the universe.

These equations, beside the geodesic equation, are the foundations
of general theory of relativity.

\section{The equivalent form}
Einstein field equations can also be written in the following
equivalent form:
$$
R_{\mu\nu}-g_{\mu\nu}\Lambda=\frac{8\pi
G}{c^4}(T_{\mu\nu}-\frac{1}{2}Tg_{\mu\nu}),
$$
which becomes useful when we are interested in weak-filed limits,
where $g_{\mu\nu}$ can be substituted by the Minkowski metric,
with an acceptable accuracy.

\section{The cosmological constant}
Historically, Einstein had included a cosmological constant in his
equation, relating to the metric.
\begin{equation}
R_{\mu\nu}-\frac{1}{2}g_{\mu\nu}R+g_{\mu\nu}\Lambda=8\pi
T_{\mu\nu}.\label{5-3}
\end{equation}
Since $\Lambda$ is a constant, the conservation of energy is still
valid. The cosmological term, initially introduced by Einstein to
describe a static universe (a universe with no expansion or
contraction). This attempt was unsuccessful because of two
reasons; first, the so-called static universe was unstable, and
second, the Hubble's observations of distant galaxies, confirmed
that our universe is not static, and is indeed expanding.
Therefore since Einstein had called $\Lambda$ as his
\textit{biggest blunder}, it was taken to be zero, during the next
decades \cite{26}.

Despite this misguiding for introducing the cosmological constant,
recent observational instruments, have assigned a definite value
for it which is confirmed by some observational data \cite{27,28}.

Einstein have considered his cosmological constant, to be an
independent variable. However, the corresponding term in the field
equation (\ref{1-3}) can be algebraically moved to the other side,
to become a part of the energy-momentum tensor.
\begin{equation}
T^{(\textmd{{vac}})}_{\mu\nu}=-\frac{\Lambda c^4}{8\pi
G}g_{\mu\nu}.\label{6-3}
\end{equation}
The vacuum energy is constant and is given by
\begin{equation}
\rho_{(\textmd{{vac}})}=\frac{\Lambda c^2}{8\pi G}.\label{7-3}
\end{equation}
Therefore the existence of the cosmological constant, is
equivalent to the existence of a nonzero vacuum energy. These
terms are currently being used in general relativity.

\section{Energy-momentum conservation}
In general relativity, the energy-momentum conservation law is
expressed in the following form:
\begin{equation}
\nabla_b T^{ab}=T^{ab}_{;b}=0.\label{8-3}
\end{equation}
This relation can be derived from Bianchi identity, which can be
summarized as follows:
\begin{equation}
R_{ab[cd;e]}=0,\label{9-3}
\end{equation}
for which, a multiplication by $g_{ab}$, knowing that the metric
tensor is a covariant constant, gives:
\begin{equation}
R^c_{bcd;e}+R^c_{bec;d}+R^c_{bde;c}=0.\label{10-3}
\end{equation}
The antisymmetric property of the Riemann tensor, provides this
opportunity to rewrite the second term of (\ref{10-3}), in the
following way:
\begin{equation}
R^c_{bcd;e}-R^c_{bce;d}+R^c_{bde;c}=0,\label{11-3}
\end{equation}
which is equivalent to
\begin{equation}
R_{bd;e}-R_{be;d}+R^c_{bde;c}=0.\label{12-3}
\end{equation}
To obtain (\ref{12-3}), we used the definition of the Ricci
tensor. Now multiplying both sides of (\ref{12-3}) to the metric
tensor, once again we make another contraction.
\begin{equation}
g^{bd}(R_{bd;e}-R_{be;d}+R^c_{bde;c})=0,\label{13-3}
\end{equation}
getting
\begin{equation}
R^d_{d;e}-R^d_{e;d}+R^{cd}_{de;c}=0.\label{14-3}
\end{equation}
Applying Riemann tensor and Ricci scalar definitions, one obtains:
\begin{equation}
R_{;e}-2R^{c}_{e;c}=0,\label{15-3}
\end{equation}
which can be rewritten as
\begin{equation}
(R^c_{e}-\frac{1}{2}g^c_eR)_{;c}=0.\label{16-3}
\end{equation}
One more, subtraction by $g^{ab}$ yields:
\begin{equation}
(R^{cd}-\frac{1}{2}g^{cd}R)_{;c}=0,\label{17-3}
\end{equation}
from which, the symmetry in the parenthesis and the definition of
Einstein tensor implies that:
\begin{equation}
G^{ab}_{;b}=0.\label{18-3}
\end{equation}
Using Einstein equations, we immediately conclude
$$\nabla_b T^{ab}=T^{ab}_{;b}=0,$$
which is the same as equation (\ref{8-3}). This relation locally
expresses the energy-momentum conservation. This conservation law
is of physical interest. In his field equations, Einstein confirms
that general relativity possesses such conservation law.

\section{Non-linearity}
The non-linearity of Einstein equations, makes general relativity
a distinguishable theory, among the others in physics. For
example, Maxwell's equations for electromagnetism, are linear in
electric and magnetic fields and corresponding distributions of
charge and current (since a linear combination of two solutions,
is itself a solution). Another example is the Schr\"{o}dinger
equation in quantum mechanics, which is linear with respect to the
wave function.

\section{The equivalence principle} Einstein field equations and
its approximations for weak fields and low speeds, lead to
Newton's law of gravitation. In fact the constant which appears in
Einstein equations, is determined from these two approximations.

\section{The vacuum field equations}
If the energy-momentum tensor $T_{\mu\nu}$ becomes zero in a
definite region, as a consequence, the field equations transform
to the vacuum case. Taking $T_{\mu\nu}=0$ in all filed equations,
one can write the vacuum field equations as follows:
\begin{equation}
R_{\mu\nu}=\frac{1}{2}Rg_{\mu\nu}.\label{19-3}
\end{equation}
Deriving the trace of this equation (contraction by $g_{\mu\nu}$)
and knowing that $g^{\mu\nu}g_{\mu\nu}=4$, we will have:
\begin{equation}
R=\frac{1}{2}R\times (4)=2R,\label{20-3}
\end{equation}
or
\begin{equation}
R=0.\label{21-3}
\end{equation}
Substituting (\ref{21-3}) in (\ref{19-3}) provides a corresponding
form for the vacuum equations.
\begin{equation}
R_{\mu\nu}=0.\label{22-3}
\end{equation}
When a nonzero cosmological constant is included, we will have:
\begin{equation}
R_{\mu\nu}=\frac{1}{2}Rg_{\mu\nu}-\Lambda g_{\mu\nu},\label{23-3}
\end{equation}
from which one obtains
\begin{equation}
R=4\Lambda.\label{24-3}
\end{equation}
This relation also has a corresponding vacuum form like
\begin{equation}
R_{\mu\nu}=\Lambda g_{\mu\nu}.\label{25-3}
\end{equation}
The solutions to these equations are called \textit{vacuum
solutions}. Minkowski spacetime is the simplest vacuum solution to
Einstein equations. Some other nontrivial solutions are the
Schwarzchild and Kerr solutions.

Manifolds with $R_{\mu\nu}=0$, are called \textit{Ricci-flat
manifolds}, and the manifolds in which Ricci tensor is
proportional to the metric, are called \textit{Einstein
manifolds}.

\section{Einstein-Maxwell equations}
If the energy-momentum tensor is considered to be
\begin{equation}
T^{\alpha\beta}=-\frac{1}{\mu_0}(F^{\alpha\beta}F^\beta_\psi+\frac{1}{4}g^{\alpha\beta}F_{\psi\tau}F^{\psi\tau}),\label{26-3}
\end{equation}
corresponding to an electromagnetic field in free space, then the
resultant equations from (\ref{1-3}), get the Einstein-Maxwell
field equations.
\begin{equation}
R^{\alpha\beta}-\frac{1}{2}Rg^{\alpha\beta}+g^{\alpha\beta}\Lambda=-\frac{8\pi
G}{c^4\mu_0}(F^{\alpha\beta}F^\beta_\psi+\frac{1}{4}g^{\alpha\beta}F_{\psi\tau}F^{\psi\tau}).\label{27-3}
\end{equation}
Furthermore, the covariant form of Maxwell's equations are also
applicable in free space.
\begin{equation}
F^{\alpha\beta}_{;\beta}=0,\label{28-3}
\end{equation}
and
\begin{equation}
F_{[\alpha\beta;\gamma]}=\frac{1}{3}(F_{\alpha\beta;\gamma}+F_{\beta\gamma;\alpha}+F_{\gamma\alpha;\beta})=0,\label{29-3}
\end{equation}
in which $_;$ denotes covariant differentiation, and the brackets
represents the antisymmetry.  The first equation asserts that a
four-dimensional divergence of $F$, vanishes.

The second equation, equates exterior differentiation to zero.
Therefore from Poincar\`{e} lemma, we can define a vector
potential $A_\alpha$ so that
\begin{equation}
F_{\alpha\beta}=A_{\alpha;\beta}-A_{\beta;\alpha}=A_{\alpha,\beta}-A_{\beta,\alpha},\label{30-3}
\end{equation}
where $_,$ stands for partial differentiations. Equation
(\ref{30-3}) sometimes is regarded as covariant Maxwell equation
\cite{29}. However, there are some generalized solutions to this
equation, without possessing a generalized definite potential
\cite{30}.

\section{The solutions}
The solutions of Einstein equations are the spacetime structures.
Hence, this solutions are named \textit{metrics}. These metrics
describe the spacetime structure, including the inertial
characteristics of the contained objects. Since the field
equations are nonlinear, it is not possible to solve them
completely (inevitably we must make approximations). For example,
there is no known complete solution for the spacetime, containing
two separated masses (which is the theoretical model for dipole
stars). However in such problems, it is common to consider some
sort of approximations, called \textit{post-Newtonian
approximations}. Although in several cases, the field equations
have been solved completely, resulting in \textit{exact solutions}
\cite{31}.

Probing the exact solutions of Einstein equations, is one of the
most important activities in cosmology. These investigations, have
led to anticipations like black-hole existence, and also to
proposing some different models of universe's evolution.

\section{The linearized field equations}
The linearized Einstein equation is an approximation, which is
valid for weak-filed limits and is applied to simplifying the
problems in general relativity or discussing the concept of
gravitational waves. This approximation can also be used to deduce
Newtonian gravitation from general relativity, as the weak-field
limit.

These approximations can be derived, by considering that the
spacetime metric, differs only slightly from the Minkowskian one.
Then the difference between the metrics can be regarded as a field
on the original metric (background metric), and its behavior is
investigated by a set of linear equations.

\subsection{Derivation of Minkowski metric}
Let us consider the following metric for our spacetime:
\begin{equation}
g_{ab}=\eta_{ab}+h_{ab},\label{31-3}
\end{equation}
where $\eta_{ab}$ is the Minkowski metric and $h_{ab}$ correlated
to a field located on the background metric. $h$ has to be
ignorable versus $\eta$. That is $|h_{\mu\nu}|\ll 1$ (and also for
all derivatives of $h$). Therefore one can ignore all the
multiplications of $h$ by itself (and all its derivatives with
respect to $h$). Then we shall assume that all the indices of $h$
can be raised or lowered by a $\eta$.

$h$ is always symmetric, and the condition
$g_{ab}g^{bc}=\delta^c_a$ implies that
\begin{equation}
g^{ab}=\eta^{ab}+h^{ab},\label{32-3}
\end{equation}
from which the Christoffel symbols can be calculated as follows:
\begin{equation}
2\Gamma^a_{bc}=(h^a_{b,c}+h^a_{c,b}-h^a_{bc,}),\label{33-3}
\end{equation}
where
$$h^a_{bc,}:=\eta^{ar}h_{bc,r}$$
is used to calculate the Riemann tensor
\begin{equation}
2R^a_{bcd}=2(\Gamma^a_{bd,c}+\Gamma^a_{bc,d})
$$$$=\eta^{ac}(h_{eb,dc}+h_{ed,bc}-h_{bd,ec}-h_{eb,cd}-h_{ec,bd}+h_{bc,ed})
$$$$=\eta^{ac}(h_{ed,bc}-h_{bd,ec}-h_{ec,bd}+h_{bc,ed})
$$$$=(h^a_{d,bc}-h^a_{bd,c}+h^a_{bc,d}-h^a_{c,bd}).           \label{34-3}
\end{equation}
Using $R_{ab}=\delta^c_a R^a_{bcd}$ we have:
\begin{equation}
2R_{bd}=h^r_{d,br}+h^r_{b,dr}-h_{,bd}-h_{bd,rs}\eta^{rs}.\label{35-3}
\end{equation}
Therefore the linearized Einstein equations take the form
\footnote{Stephani, Hans (1990), \emph{General Relativity: An
Introduction to the Theory of Gravitation Filed}, Cambridge
University Press. ISBN 0-521-37941-5.}:
\begin{equation}
8\pi
T_{ab}=R_{bd}-\frac{1}{2}R_{ac}\eta^{ac}\eta_{bd}.\label{36-3}
\end{equation}

\section{The de Sitter space}
In mathematical physics, the $n$-dimensional de Sitter space,
denoted by $dS_n$, is the Lorentz analogy of a $n$-dimensional
sphere (with a canonic metric). This space is a maximally
symmetric Lorentz manifold, with a constant positive curvature,
which would be continuous for $n\geq3$. From general relativity
view point, the de Sitter space is a maximally symmetric vacuum
solution to Einstein equations, possessing a positive cosmological
constant (repelling) corresponding to a positive density of vacuum
energy with negative pressure. When $n=4$, de Sitter space is also
a cosmological model. This space has been proposed by Willem de
Sitter, and independently by Tullio Levi-Civita in 1917.

Formerly, this space was considered as a basis of general
relativity, instead of Minkowski space, forming a formalism called
\textit{de Sitter relativity} \cite{32}.

\subsection{Mathematical definition}
De Sitter space is a sub-manifold of Minkowski space with one
additional dimension. Let us consider a Minkowski space $R^{1,n}$
with the standard metric
\begin{equation}
ds^2=-dx_0^2+\sum_{i=1}^n dx_i^2.\label{37-3}
\end{equation}
The so-called sub-manifold is a hyperbolic surface, defined by
\begin{equation}
-x_0^2+\sum_{i=1}^n x_i^2=\alpha^2,\label{38-3}
\end{equation}
in which $alpha$ is a positive constant having the dimension of
length. The induced metric which is defined on de Sitter space, is
deduced from Minkowski ambient space. One can inspect that this
induced metric, is non-degenerate exhibits a Lorentz form. Note
that, if $\alpha^2$ is substituted by $-\alpha^2$ in the above
definition, a two-surfaced hyperboloid is achieved. The so-called
induced metric in this case, is definite and positive, and every
surface, is a copy of a $n$-dimensional hyperboloid.

De Sitter space can also be regarded as the quotient of the
fraction $\frac{O(1,n)}{O(1,n-1)}$ of two indefinite nonorthogonal
groups, which shows that this space is symmetric and
non-Riemannian. From topological view point, de Sitter space is
$\mathbb{R}\times s^{n-1}$, therefore for $n\geq3$, is continuous
\cite{33,34,35}.

\subsection{Properties}
The isometry group od de Sitter space, is the Lorentz group
$O(1,n)$. Therefore its metric is maximally symmetric and has
$\frac{n(n+1)}{2}$ independent Killing vectors. Every maximally
symmetric spaces, have a constant curvature. The corresponding
Riemann tensor is given by
\begin{equation}
R_{\rho\mu\nu\lambda}=\frac{1}{\alpha^2}(g_{\mu\nu}g_{\sigma\nu}-g_{\rho\nu}g_{\sigma\mu}).\label{39-3}
\end{equation}
Since the Ricci tensor is proportional to the metric, de Sitter
space is an Einstein manifold.
\begin{equation}
R_{\mu\nu}=\frac{n-1}{\alpha^2}g_{\mu\nu}.\label{40-3}
\end{equation}
This means that de Sitter space is a vacuum solution to Einstein
equations, with the cosmological constant
\begin{equation}
\Lambda=\frac{(n-1)(n-2)}{2\alpha^2}.\label{41-3}
\end{equation}
The constant curvature of de Sitter space is given by the
following relation:
\begin{equation}
R=\frac{n(n-1)}{\alpha^2}=\frac{2n}{n-2}\Lambda.\label{42-3}
\end{equation}
For $\Lambda=4$ we have $\Lambda=\frac{3}{\alpha^2}$ and
$R=4\alpha$ \cite{36,37}.

\section{De Sitter group and invariance}
The de Sitter group $SO_0(1,4)$ is composed of all $g_{5\times5}$
matrices with determinant 1, preserving the following quadratic
form:
$$x_0^2-x_1^2-x_2^2-x_3^2-x_4^2.$$
The compact maximally symmetric subgroup $k$ of $SO_0(1,4)$ is
isometric to $SO(4)$ and is composed of matrices
$$\left| {\begin{array}{*{20}{c}}
  {{1}}&{0}\\
  {0}&{k} \\
\end{array}} \right|,$$
and $k\in SO(4)$ \cite{14,15,23}.

\section{The coordinate systems in de Sitter spacetime}

\subsection{Static coordinate system}
It is possible to define the static coordinates $(t,r,...)$ in de
Sitter space as follows:
$$x_0=\sqrt{\alpha^2-r^2}\sinh(\frac{t}{\alpha}),$$
$$x_1=\sqrt{\alpha^2-r^2}\cosh(\frac{t}{\alpha}),$$
$$x_i=rz_i\,\,\,\,\,\,\,\,\,\,\,\,\,\,\,\,\,\,2\leq i\leq n,$$
where $z_i$ is the contained space in a $(n-2)$-dimensional sphere
in $\mathbb{R}^{n-1}$. In this system, the de Sitter metric has
the form
\begin{equation}
ds^2=-(1-\frac{r^2}{\alpha^2})dt^2+(1-\frac{r^2}{\alpha^2})^{-1}dr^2+r^2
d\Omega_{n-2}^2.\label{43-3}
\end{equation}
This metric is also known as the Einstein-de Sitter metric. Note
that in this coordinate system, there would be an \textit{event
horizon} (cosmological horizon) at $r=\alpha$, as a result of the
singularity in this point.

\subsection{Flat coordinate system}
If we have
$$x_0=\alpha\sinh(\frac{t}{\alpha})+\frac{r^2}{2\alpha}e^{\frac{t}{\alpha}},$$
$$x_1=\alpha\cosh(\frac{t}{\alpha})-\frac{r^2}{2\alpha}e^{\frac{t}{\alpha}},$$
$$x_i=e^{\frac{t}{\alpha}}y_i\,\,\,\,\,\,\,\,\,\,\,\,\,\,\,\,\,\,2\leq i\leq n,$$
where $r^2=\sum_i y_i^2$, therefore for this system, the de Sitter
metric in $(t,y_i)$ coordinates will be
\begin{equation}
ds^2=-dt^2+e^{\frac{2t}{\alpha}}dy^2,\label{44-3}
\end{equation}
in which $dy^2=\sum_i dy_i^2$ is the flat metric on $y_i$.

\subsection{Open slicing}
In this system we have
$$x_0=\alpha\sinh(\frac{t}{\alpha})\cosh(\xi),$$
$$x_1=\alpha\cosh(\frac{t}{\alpha}),$$
$$x_i=\alpha z_i\sinh(\frac{t}{\alpha})\sinh(\xi)\,\,\,\,\,\,\,\,\,\,\,\,\,\,\,\,\,\,2\leq i\leq n,$$
where $\sum_i z_i^2=1$, forms a $s^{n-1}$ with the standard metric
$\sum_i dz_i^2=d\Omega^2_{n-2}$. In this case the de Sitter metric
takes the form
\begin{equation}
ds^2=-dt^2+\alpha^2\sinh(\frac{t}{\alpha})dH^2_{n-1},\label{45-3}
\end{equation}
where
$$dH_{n-1}^2=d\xi^2+\sinh^2(\xi) d\Omega^2_{n-1}$$
is a hyperbolic Euclidean space.

\subsection{Closed slicing}
Assume that
$$x_0=\alpha\sinh(\frac{t}{\alpha}),$$
$$x_i=\alpha \cosh(\frac{t}{\alpha})z_i\,\,\,\,\,\,\,\,\,\,\,\,\,\,\,\,\,\,1\leq i\leq n,$$
where $z_i$ describes a $s^{n-1}$. Therefore the metric will be
\begin{equation}
ds^2=-dt^2+\alpha^2\cosh^2(\frac{t}{\alpha})d\Omega^2_{n-1}.\label{46-3}
\end{equation}
Changing the time variable to the conformal time using
$\tan(\frac{\eta}{2})=\tanh(\frac{t}{2\alpha})$ (or equivalently
$\cos(\eta)=\frac{t}{\cosh(\frac{t}{\alpha})}$), we obtain a
metric, which is conformally equivalent to a static Einstein
universe.
\begin{equation}
ds^2=\frac{\alpha^2}{\cos^2(\eta)}(-d\eta^2+d\Omega^2_{n-1}).\label{47-3}
\end{equation}
This metric is of benefit when we concern about the Penrose
diagrams of de Sitter spacetime.

\subsection{De Sitter slicing}
If we have
$$x_0=\alpha\sin(\frac{\chi}{\alpha})\sinh(\frac{t}{\alpha})\cosh(\xi),$$
$$x_1=\alpha\cos(\frac{\chi}{\alpha}),$$
$$x_2=\alpha\sin(\frac{\chi}{\alpha})\cosh(\frac{t}{\alpha}),$$
$$x_i=\alpha z_i\sin(\frac{\chi}{\alpha})\sinh(\frac{t}{\alpha})\sinh(\xi)\,\,\,\,\,\,\,\,\,\,\,\,\,\,\,\,\,\,3\leq i\leq n,$$
where $z_i$ describes a $s^{n-1}$, the metric takes the form
\begin{equation}
ds^2=d\chi^2+\sin^2(\frac{\chi}{\alpha})ds^2_{\textmd{\tiny{ds}},\alpha,n-1},\label{48-3}
\end{equation}
in which
$$ds^2_{\textmd{\tiny{ds}},\alpha,n-1}=-dt^2+\alpha^2\sinh^2(\frac{t}{\alpha})dH^2_{n-2}$$
is the metric of a $(n-1)$-dimensional de Sitter space with
$\alpha$ as the radius of curvature in open coordinate system. The
hyperbolic metric is given by the following relation:
\begin{equation}
dH_{n-2}^2=d\xi^2+\sinh^2(\xi) d\Omega_{n-3}^2.\label{49-3}
\end{equation}
Continuing the open coordinate system analysis, this metric is
derived from the transformation
$$(t,\xi,\theta,\phi_1,\phi_2,...,\phi_{n-3})\longrightarrow(i\chi,\xi,it,\phi_2,...,\phi_{n-4}),$$
and also the substitution $x_0$ by $x_2$, since these two, replace
their timelike/spacelike natures \cite{38}.

\section{Solutions to Einstein equations from Birkhoff's view point}
The Schwarzchild metric
\begin{equation}
ds^2=-(1-\frac{2GM}{r})dt^2+(1-\frac{2GM}{r})^{-1}dr^2+r^2d\theta^2+r^2\sin^2\theta
d\phi^2, \label{Schwarzchild}
\end{equation}
is known to be directly related to the spherically symmetric
metric, as the original result of the Birkhoff's theorem in 1923
\cite{Goenner}. This metric is the vacuum solution to Einstein
field equations, explaining the exterior geometry of a spherical
object of mass $M$. One can consider the Birkhoff's theorem as
follows:\\

{\textit{A spcetime is spherically symmetric, if there exists a
$SO(3)$ group of isometries on it, which are isomorphic to a
$S^2$-sphere.}}\\

Nevertheless, the Schwarzchild solution is supposed to be static,
the results of Birkhoff's theorem is not confined to static
spherically symmetric metrics. According to the previous section,
the Einstein-de Sitter metric
\begin{equation}
ds^2=-(1-\frac{2GM}{r}-\frac{1}{3}\Lambda
r^2)dt^2+(1-\frac{2GM}{r}-\frac{1}{3}\Lambda
r^2)^{-1}dr^2+r^2d\Omega^2 \label{E-d}
\end{equation}
where $\Lambda$ is the cosmological constant, is also capable to
be considered as a result of Birkhoff's theorem. This metric has
to be the trivial solution to vacuum Bach's conformal field
equations, derived from conformal Weyl gravity \cite{Mannheim},
i.e.
\begin{equation}
W_{\alpha\beta}=\nabla^\rho\nabla_\alpha
R_{\beta\rho}+\nabla^\rho\nabla_\beta R_{\alpha\rho}-\Box
R_{\alpha\beta}-g_{\alpha\beta}\nabla_\rho\nabla_\lambda
R^{\rho\lambda}$$$$ -2R_{\rho\beta}
R^{\rho}_\alpha+\frac{1}{2}g_{\alpha\beta}R_{\rho\lambda}R^{\rho\lambda}-\frac{1}{3}\Big(2\nabla_\alpha\nabla_\beta
R-2g_{\alpha\beta}\Box
R-2RR_{\alpha\beta}+\frac{1}{2}g_{\alpha\beta}R^2\Big)=0.
\label{bach}
\end{equation}
This holds, since de Sitter spacetime is conformally flat and
obeys the $SO(d,2)$ conformal algebra, and therefore it would be
conformally symmetric. What we are about to do in this work, is to
derive conformally invariant equations for massless particles with
spin-2, on a de Sitter background. Previously, the conformal
invariance of such particles, beside investigating this property
for scalars and partially massless particles in de Sitter and
Anti-de Sitter spacetimes, have been studied \cite{Deser}. In this
work however, we shall use a different approach, namely the
Dirac's six cone formalism to reconsider this, which will be
introduced in the next chapter.

\chapter{The mathematical operators and Dirac's six cone formalism}
\section{Introducing the transverse projector}
Lets consider a spherical surface in $3$-dimensional space.
Cartesian expression of this surface would be
$$x^2+y^2+z^2=R^2=1,$$
indicating a spherical shell with a unit radius. Now let us
introduce the transverse projector operator on this shell, in
three-dimensional space as
\begin{equation}
\theta^{ij}A_j = \bar A^i \label{1-4}
\end{equation}
where $\bar A^i$ are the components of $\overrightarrow{A}$ on the
sphere's surface. Since $\bar A^i$ is a tangent vector, it is
perpendicular to the radius vector of the sphere, so
\begin{equation}
r_i\theta^{ij}A_j=r_i\bar A^i=0. \label{2-4}
\end{equation}
generally this operator can be defined as below:
\begin{equation}
\theta_{ij}=\delta_{ij}-r^{-2}r_ir_j. \label{3-4}
\end{equation}
Corresponding to the operator defined in (\ref{1-4}), we introduce
a transverse projector, to project a vector from de Sitter ambient
(flat) space to de Sitter inherent (curved-hyperbolic) space.
\begin{equation}
\theta_{\alpha\beta}=\eta_{\alpha\beta}+H^2x_\alpha x_\beta,
\label{4-4}
\end{equation}
in which $\alpha,\beta=1,2,3,4,5$ and
$$\eta_{\alpha\beta}=diag(1,-1,-1,-1,-1).$$
The metric will be
$$ds^2=\eta_{\alpha\beta}dx^\alpha dx^\beta=g^{\textmd{\small{ds}}}_{\mu\nu}dX^\mu dX^\nu.$$
It can be shown that
\begin{equation}
x_\alpha\bar A^\alpha=x^\alpha\bar A_\alpha=0. \label{5-4}
\end{equation}
Note that
\begin{equation}
x_\alpha x^\alpha=-H^{-2} \label{6-4}
\end{equation}
which is deduced from de Sitter space definition, and is
equivalent to the following relation:
\begin{equation}
x_0^2-x_1^2-x_2^2-x_3^2-x_4^2=-H^{-2}. \label{7-4}
\end{equation}
The differentiation operators are also capable to be projected
from ambient space onto curved space. In general relativity, this
kind of differentiation is called covariant differentiation. We
have
\begin{equation}
\bar\partial_\alpha =
\theta_{\alpha\beta}\partial^\beta=\partial_\alpha+H^2x_\alpha(x.\partial).
\label{8-4}
\end{equation}
One can easily show that $\bar\partial_\alpha$ is the transverse
component of differentiation in the curved space, i.e.
\begin{equation}
x_\alpha\bar\partial^\alpha=x^\alpha\bar\partial_\alpha=0.
\label{9-4}
\end{equation}
Note that
\begin{equation}
\frac{\partial x^\alpha}{\partial
x^\beta}=\delta^\alpha_\beta=\eta^\alpha_\beta \label{10-4}
\end{equation}
and
\begin{equation}
\bar\partial^\alpha x_\beta=
\theta^\alpha_\beta\,\,\,\,\,\Rightarrow\,\,\,\,\,\bar\partial^\alpha
x_\alpha=4. \label{11-4}
\end{equation}
It can be shown that for the Hubble parameter we have
\begin{equation}
\bar\partial_\alpha H^{-2}=0, \label{12-4}
\end{equation}
and
\begin{equation}
\partial_\alpha H^{-2}=-2x_\alpha.
\label{13-4}
\end{equation}
The vector components can be differentiated in the following way:
\begin{equation}
\bar\partial_\alpha x^\beta=\theta^\beta_\alpha. \label{14-4}
\end{equation}
Note that
\begin{equation}
\partial_\gamma x^\beta=\eta^\beta_\gamma.
\label{15-4}
\end{equation}
Therefore it can proved that
\begin{equation}
\bar\partial_\beta x^\beta=4. \label{16-4}
\end{equation}

\section{Applying the transverse projector on operators in de
Sitter spacetime} Defining
$$M_{\alpha\beta}=i(x_\alpha\partial_\beta-x_\beta\partial_\alpha),$$
It can be shown that
\begin{equation}
(x_\alpha\partial_\beta-x_\beta\partial_\alpha)(x^\alpha\partial^\beta-x^\beta\partial^\alpha)
$$$$=(x_\alpha\bar\partial_\beta-x_\beta\bar\partial_\alpha)(x^\alpha\bar\partial^\beta-x^\beta\bar\partial^\alpha).\label{17-4}\end{equation}
The scalar Casimir operator of de Sitter group is defined as
\begin{equation}
Q_0=-\frac{1}{2}M_{\alpha\beta}M^{\alpha\beta}, \label{18-4}
\end{equation}
from which
\begin{equation}
Q_0=H^{-2}\bar\partial_\alpha\bar\partial_\alpha. \label{19-4}
\end{equation}
Applying the transverse projector on itself, yields
\begin{equation}
\theta_{\alpha\beta}\theta^{\alpha\beta}=\theta_\beta^\gamma.
\label{20-4}
\end{equation}
It is also possible to calculate covariant differentiations of the
transverse projector.
\begin{equation}
\bar\partial_\beta\theta_{\alpha\sigma}=H^2\theta_{\beta\alpha}x_\sigma+H^2
x_\alpha\theta_{\beta\sigma}. \label{21-4}
\end{equation}
For an arbitrary function $f$, we have
\begin{equation}
\bar\partial_\beta(x_\gamma
f)=\theta_{\beta\gamma}f+x_\gamma\bar\partial_\beta f.
\label{22-4}
\end{equation}
One can prove that
\begin{equation}
Q_0 x^\beta=-4x^\beta. \label{23-4}
\end{equation}
Differentiation in curved space, is not commutative and indeed has
the following cummutation relation:
\begin{equation}
[\bar\partial_\alpha,\bar\partial_\beta]=H^2(x_\beta\bar\partial_\alpha-
x_\alpha\bar\partial_\beta). \label{24-4}
\end{equation}
Also some other cummutation relations are valid and provable.
\begin{equation}
[Q_0,x_\alpha]=-4x_\alpha-2H^{-2}\bar\partial_\alpha, \label{25-4}
\end{equation}
\begin{equation}
[Q_0,\bar\partial_\alpha]=
6\bar\partial_\alpha+2H^2(Q_0+4)x_\alpha. \label{26-4}
\end{equation}
We should mention that
\begin{equation}
x_\alpha\theta_{\alpha\gamma}=0. \label{27-4}
\end{equation}

\section{The differentiation operator in de Sitter spacetime}
The differentiation operator is defined like
\begin{equation}
D_{1\alpha}\equiv H^{-2}\bar\partial_\alpha, \label{28-4}
\end{equation}
or
\begin{equation}
D_1\equiv H^{-2}\bar\partial.\label{29-4}
\end{equation}
It is proved that
\begin{equation}
Q_1D_1k_\alpha=D_1Q_0\phi,\label{30-4}
\end{equation}
in which $\phi$ is an arbitrary scalar field like the formerly
introduced function $f$. Note that
\begin{equation}
Q_1 k_\alpha = (Q_0-2)k_\alpha+2
x_\alpha\bar\partial.k,\label{31-4}
\end{equation}
where $k_\alpha$ is a vector field. Therefore in general, for a
vector field we have
\begin{equation}
Q_1 k_\alpha = (Q_0-2)k_\alpha + 2x_\alpha\bar\partial.k -
2\bar\partial x.k.\label{32-4}
\end{equation}
And since $k$ is supposed to be transverse, therefore
\begin{equation}
x.k=0.\label{33-4}
\end{equation}
The Casimir operators are classified as
\begin{itemize}
\item{$Q_0$ operates on scalars,} \item{$Q_1$ operates on
vectors,} \item{$Q_2$ operates on tensors (two components).}
\end{itemize}

If $\bar z_\alpha$ is an arbitrary constant vector, its
derivatives are zero, however note that
\begin{equation}
\bar z_\alpha = \theta_{\alpha}^{\beta} z_\beta.\label{34-4}
\end{equation}
That is because of the existence of $\theta^\beta_\alpha$ its
derivatives do not vanish.
\begin{equation}
\partial . \bar z\phi = \partial^\alpha\bar z_\alpha\phi = \partial^\alpha\theta_{\alpha\beta} z^\beta\phi
$$$$=4H^2(x.z)\phi+z^\beta\bar\partial_\beta\phi=4H^2(x.z)\phi+z.\bar\partial\phi.\label{35-4}
\end{equation}
Also we have
\begin{equation}
Q_0(x.z)\phi=(x.z)(Q_0-4)\phi+2H^{-2}z.\bar\partial\phi.\label{36-4}
\end{equation}
One can prove
\begin{equation}
Q_0(z.\bar\partial)\phi=z.\bar\partial(Q_0+1)\phi+2H^2(x.z)Q_0\phi.\label{37-4}
\end{equation}

\section{Transversifying the vectors on de Sitter spacetime }
The Laplace-Belzami operator is defined as
\begin{equation}
\Box\equiv -H^{-2}Q_0.\label{38-4}
\end{equation}
Since we are now familiar with the basics concepts of
$\theta_{\alpha\beta}$, $Q_0$, $Q_1$ and $D_1$ and their
commutation relations, it is worth to know that the transposition
of a differentiation between de Sitter ambient space and coherent
space, is done in the following way:
\begin{equation}
\nabla_\mu A_\nu \longrightarrow
\theta_\alpha^{\alpha^\prime}\theta_\beta^{\beta^\prime}\partial_{\alpha^\prime}k_{\beta^\prime}.\label{39-4}
\end{equation}
Since we possessed two indices, we had to differentiate twice.
This is called \textit{transverse projection}, which is noted by
$\textmd{TrPr}$. We have:
\begin{equation}
\nabla_\rho\nabla_\lambda h_{\mu\nu} \longrightarrow
(\textmd{TrPr})\partial_\alpha(\textmd{TrPr})\partial_\beta
k_{\mu\nu}.\label{40-4}
\end{equation}
The term $\textmd{TrPr}(\partial_\beta k_{\mu\nu})$ must results
in an expressions, which vanishes when multiplied by an $x$; in
other words the result must be transverse.
\begin{equation}
\textmd{TrPr}(\partial_\beta k_{\mu\nu}) = \bar\partial_\beta
k_{\mu\nu} - x_\mu k_{\beta\nu} - x_\nu k_{\mu\beta}.\label{41-4}
\end{equation}
Since $k$ is transverse itself ($x.k=0$), the rhs of (\ref{41-4})
results in zero when it is multiplied by $x_\beta$, $x_\mu$ or
$x_\nu$.
\begin{equation}
x_\beta \textmd{TrPr}(\partial_\beta k_{\mu\nu})=x_\mu
\textmd{TrPr}(\partial_\beta k_{\mu\nu}) = x_\nu
\textmd{TrPr}(\partial_\mu k_{\mu\nu}) = 0.\label{42-4}
\end{equation}
The other calculations for $\textmd{TrPr}$, can be done in the
same way. For example it can be proved that
\begin{equation}
\nabla_\mu A_\nu \longrightarrow \textmd{TrPr}(\partial_\alpha
k_\beta) = \bar\partial_\alpha k_\beta - H^2 x_\beta
k_\alpha.\label{43-4}
\end{equation}
Note that it is common to consider $H^2=1$.

\section{Dirac's six cone formalism}
The Dirac's six cone is a procedure through which the conformally
invariant equations are derived in six-dimensional space. Here the
variables are denoted by $u^a$, where:
$$u^a\,\,\,\,\,\,\,\,\,\,\,\,\,\,\,\,a=0,1,2,3,4,5.$$
From differential calculus, we know that $x.\bar\partial$
indicates the degree of a differentiable function.  For example
for $f(x)=x^5$, this operator results in $5$. i.e.
\begin{equation}
x.\partial f(x)= 5 f(x).\label{44-4}
\end{equation}
Similarly in Dirac's six cone formalism, the operator
$u^a\partial_a$ is regarded as the conformal degree operator. The
filed is denoted by $\psi$ and the following relation is applied
to find the equations in six-dimensional space:
\begin{equation}
\left\{
\begin{array}{c}
N_5\psi=(p-2)\psi\\\\
(\partial_a\partial^a)\psi=0
\end{array}
 \right.,
 \label{45-4}
\end{equation}
where $N_5=u_a\partial^a$ and the whole set are conformally
invariant. The relation between $\partial_a\partial^a$ and the
Casimir operators of de Sitter group is
\begin{equation}
(\partial_a\partial^a)^p = -x_5^{-2p} \prod_{j=1}^p \Big(Q_0 +
(j+1)(j-2)\Big).\label{46-4}
\end{equation}
As an example, for the simplest case of $p=1$ we have
\begin{equation}
(\partial_a\partial^a)^1 = -\frac{1}{x_5^2}(Q_0 - 2),\label{47-4}
\end{equation}
where $Q_0$ is the Casimir operator of de Sitter group. Also for
$p=2$:
\begin{equation}
(\partial_a\partial^a)^2= -\frac{1}{x_5^4}(Q_0 -
2)Q_0.\label{48-4}
\end{equation}
The following relation can be regarded as the connection between
six-dimensional space and de Sitter ambient space:
\begin{equation}
\left\{
\begin{array}{c}
x^\alpha = (u^5)^{-1}u^\alpha,\,\,\,\,\,\,\,\,\,\,\, \alpha = 0,1,2,3,4\\\\
x^5 = u^5
\end{array}
 \right..
 \label{49-4}
\end{equation}
Note that $x^5$ is an extra component; it is not one of the 5
components of de Sitter ambient space and will vanish in our final
equations.

For the scalar field $\psi$ (the simplest case), the conformally
invariant system is
\begin{equation}
\left\{
\begin{array}{c}
(\partial_a\partial^a)\psi = 0\\\\
N_5\psi = -\psi
\end{array}
 \right..
 \label{50-4}
\end{equation}
Now since $\psi$ does not possess indices, and is invariant in all
systems, defining $\phi=x_5\phi$, we can construct a de Sitter
scalar in the form
\begin{equation}
(Q_0 - 2)\phi = 0.\label{51-4}
\end{equation}
This solution is valid for a massless scalar field in de Sitter
space.\\

The second stage, is a vector field for which $p=1$. We have
\begin{equation}
\left\{
\begin{array}{c}
(\partial_a\partial^a)\psi_a = 0\\\\
N_5\psi_a = -\psi_a
\end{array}
 \right..
 \label{52-4}
\end{equation}
$\psi_a$ possesses six components; it has six degrees of freedom.
Now $k_\alpha$, the vector filed in de Sitter ambient space, must
be deduced from $\psi_\alpha$ such that it is transverse.
\begin{equation}
k_\alpha = x_5(\psi_\alpha + x_\alpha x . \psi).\label{53-4}
\end{equation}
The vector in (\ref{53-4}) would be transverse. By applying the
operator defined in (\ref{51-4}) on the vector $k_\alpha$ we get
\cite{39,40}:
\begin{equation}
(Q_0 - 2)(\psi_\alpha + x_\alpha x . \psi) = 0.\label{54-4}
\end{equation}
The next step, is to derive an equation for $k_{\alpha\beta}$,
which is the last part of this thesis and we shall concern about
in the next chapter.

\chapter{The conformally invariant equations for graviton}
\section{The conformally invariant system of conformal degree 1 }
Working with $p=1$, the formalism of Dirac's six becomes
\begin{equation}
\left\{
\begin{array}{c}
(\partial_A\partial^A)\psi_{AB} = 0\\\\
N_5\psi_{AB} = -\psi_{AB}
\end{array}
 \right..
 \label{1-5}
\end{equation}
We introduce the second rank tensor $k_{\alpha\beta}$ from
$\psi_{\alpha\beta}$ in de Sitter ambient space.
\begin{equation}
k_{\alpha\beta}=\psi_{\alpha\beta}+Sx_\beta\psi_\alpha. x+x_\alpha
x_\beta x.\psi. x, \label{2-5}
\end{equation}
where $S$ denotes that the next expression is added to itself with
a commutation on its coefficients. Also the differentiations are
done like
\begin{equation}
\bar\partial.k_\alpha=3(x.\psi_\alpha+x_\alpha x.\psi.x).
\label{3-5}
\end{equation}
Now let $\partial_A\partial^A$ operate on the tensor defined in
(\ref{2-5}). From (\ref{51-4}) we know that $\partial_A\partial^A$
is equivalent to $(Q_0-2)$.We have
\begin{equation}
(Q_0-2)k_{\alpha\beta}=(Q_0-2)\psi_{\alpha\beta}+(Q_0-2)Sx_\beta\psi_\alpha.x+(Q_0-2)x_\alpha
x_\beta x.\psi.x. \label{4-5}
\end{equation}
We rewrite the conformally invariant system as
\begin{equation}
\left\{
\begin{array}{c}
(Q_0-2)\psi_{\alpha\beta} = 0\\\\
(Q_0-2)\psi_{55},\psi_{\alpha\alpha} =0
\end{array}
 \right..
 \label{5-5}
\end{equation}
The transversality condition implies that
\begin{equation}
u^A\psi_{AB}=0\,\,\,\,\,\,\,\,\,\,\Rightarrow\,\,\,\,\,\,\,\,\,\,(Q_0-2)x.\psi_\beta=0.\label{6-5}
\end{equation}
Since $k$ is traceless, therefore
\begin{equation}
k'=0\,\,\,\,\,\,\,\,\,\,\Rightarrow\,\,\,\,\,\,\,\,\,\,(Q_0-2)x.\psi.x=0\label{7-5}
\end{equation}

\section{The effect of Casimir operator on a tensor of second rank}
Multiplying a $x_\beta$ to both sides of (\ref{6-5}) yields
$$Q_0x.\psi.x+2x.\psi.x+2\bar\partial.\psi.x=0.$$
Using (\ref{7-5}) we obtain
\begin{equation}
\bar\partial.\psi.x=-2x.\psi.x.\label{8-5}
\end{equation}
Substituting (\ref{7-5}) in (\ref{8-5}) we have
\begin{equation}
(Q_0-2)\bar\partial.\psi.x=0.\label{9-5}
\end{equation}
In chapter 4 we mentioned that
$$(Q_0-2)x_\alpha=x_\alpha Q_0-6x_\alpha-2\bar\partial_\alpha,$$
hence, from (\ref{6-5}),
\begin{equation}
(Q_0-2)x_\alpha\psi_\beta.x=-2(\bar\partial_\alpha+2x_\alpha)\psi_\beta.x.\label{10-5}
\end{equation}
Therefore the effect of $(Q_0-2)$ on  a tensor of second rank, can
be summarized as follows:
\begin{equation}
(Q_0-2)k_{\alpha\beta}=-2(\bar\partial_\alpha+2x_\alpha)\psi_\beta.x-2(\bar\partial_\beta+2x_\beta)\psi_\alpha.x-2(\bar\partial_\alpha+2x_\alpha)x_\beta
x.\psi.x.\label{11-5}
\end{equation}
Using (\ref{3-5}), this can be rewritten as
\begin{equation}
(Q_0-2)k_{\alpha\beta}=-2(\bar\partial_\beta+2x_\beta)\psi_\alpha.x-\frac{2}{3}(\bar\partial_\alpha+2x_\alpha)\bar\partial.k_\beta.\label{12-5}
\end{equation}

\section{Obtaining the confomally invariant field equation using a
conformal system of degree 1} From (\ref{5-5}) and (\ref{6-5}) we
have
\begin{equation}
(Q_0-2)\psi_{\alpha\beta}=0\,\,\,\,\,\,\,\,\,\,\Rightarrow\,\,\,\,\,\,\,\,\,\,2x.\psi_\beta=-\bar\partial.\psi_\beta,\label{13-5}
\end{equation}
\begin{equation}
(Q_0-2)\psi_{\beta}.x=0\,\,\,\,\,\,\,\,\,\,\Rightarrow\,\,\,\,\,\,\,\,\,\,2x.\psi.x=-\bar\partial.\psi.x.\label{14-5}
\end{equation}
Let us multiply (\ref{4-5}) by $x_\alpha$.
\begin{equation}
2\bar\partial.k_\beta=Q_0x.\psi_\beta-2x.\psi_\beta+S\psi_\beta.x+Q_0x_\beta
x.\psi.x+2x_\beta x.\psi.x
$$$$+2\bar\partial_\beta(x_\beta x.\psi_\alpha)-Q_0 x_\beta x.\psi.x-2x_\beta x.\psi.x+8x_\beta x.\psi.x.\label{15-5}
\end{equation}
Now multiply (\ref{4-5}) by $\bar\partial_\alpha$. We have
\begin{equation}
-\frac{2}{3}Q_0\bar\partial.k_\beta+2\bar\partial.k_\beta=4\psi_\beta.x.\label{16-5}
\end{equation}
If (\ref{12-5}) is multiplied by $x_\beta$, we obtain
\begin{equation}
\bar\partial.k_\alpha=3\psi_\alpha.x,\label{17-5}
\end{equation}
or
\begin{equation}
\psi_\alpha.x=\frac{1}{3}\bar\partial.k_\alpha.\label{18-5}
\end{equation}
Substitute this equation in (\ref{12-5}).
\begin{equation}
(Q_0-2)k_{\alpha\beta}=-\frac{2}{3}(\bar\partial_\beta+2x_\beta)\bar\partial.k_\alpha
-\frac{2}{3}(\bar\partial_\alpha+2x_\alpha)\bar\partial.k_\beta.\label{19-5}
\end{equation}
Also from (\ref{16-5}) and (\ref{18-5}) we have
$$(Q_0-1)\bar\partial.k_\beta=0,$$
from which we obtain the conformally invariant system of degree 1,
for tensorial field.
\begin{equation}
(Q_0-2)k_{\alpha\beta}=-\frac{2}{3}S(\bar\partial_\beta+2x_\beta)\bar\partial.k_\alpha,\label{20-5}
\end{equation}
and
\begin{equation}
(Q_0-1)\bar\partial.k_\beta=0.\label{21-5}
\end{equation}
Nevertheless Equation (\ref{20-5}) is conformal, it does not
possess physical descriptions, since it is not capable to be
transformed by the irreducible representations of conformal group.
This equation was firstly proposed by Barut and Xu in 1982 by
varying the coefficient in the standard equation \footnote{A.O.
Barut, B.W. Xu, J. Phys. A: 15 (1982) 207.}. This equation in de
Sitter inherent space has the following form:
\begin{equation}
(\Box+4)h_{\mu\nu}=-\frac{2}{3}S\nabla_\mu\nabla.h_\nu=0.\label{22-5}
\end{equation}

\section{The conformally invariant system of conformal degree 2}
Generally we can write
\begin{equation}
\bar\partial.k_\alpha=4(\psi_\alpha.x+x_\alpha
x.\psi.x),\label{23-5}
\end{equation}
and
\begin{equation}
(Q_2+6)k_{\alpha\beta}=Q_0k_{\alpha\beta}+2Sx_\alpha\bar\partial.k_\beta+2\eta_{\alpha\beta}k'-2S\bar\partial_\alpha
x.k_\beta.\label{24-5}
\end{equation}
If $p=2$, the formalism of Dirac's six cone becomes
\begin{equation}
\left\{
\begin{array}{c}
(Q_0-2)Q_0\psi_{AB} = 0\\\\
N_5\psi_{AB} =0
\end{array}
 \right..
 \label{25-5}
\end{equation}
Let us rewrite our information.
\begin{equation}
(Q_0-2)Q_0x.\psi.x=0,\label{26-5}
\end{equation}
\begin{equation}
(Q_0-2)Q_0\psi_{\alpha\beta}=0,\label{27-5}
\end{equation}
\begin{equation}
(Q_0-2)Q_0\psi_{55}=0,\label{28-5}
\end{equation}
\begin{equation}
(Q_0-2)Q_0x.\psi_{B}=0,\label{29-5}
\end{equation}
\begin{equation}
(Q_0-2)Q_0\bar\partial.\psi.x=0.\label{30-5}
\end{equation}
The effect of $(Q_0-2)Q_0$ on the tensor defined in (\ref{2-5}) is
\begin{equation}
(Q_0-2)Q_0k_{\alpha\beta}=(Q_0-2)Q_0Sx_\alpha\psi_\beta.x+(Q_0-2)Q_0x_\alpha
x_\beta x.\psi.x.\label{31-5}
\end{equation}

\section{Obtaining the confomally invariant field equation using
a conformal system of degree 2} It can be shown that
\begin{equation}
x_\alpha(Q_0-2)Q_0=(Q_0-2)(Q_0x_\alpha+4x_\alpha+4\bar\partial_\beta).\label{32-5}
\end{equation}
Also we have
\begin{equation}
(Q_0-2)Q_0x_\alpha\psi_\beta.x=x_\alpha(Q_0-2)Q_0\psi_\beta.x-4(3x_\alpha+\bar\partial_\alpha)(Q_0-2)x_\beta
x.\psi.x,\label{33-5}
\end{equation}
and
\begin{equation}
(Q_0-2)Q_0x_\alpha x_\beta x.\psi.x=x_\alpha x_\beta(Q_0-2)Q_0
x.\psi.x$$$$-4x_\alpha(3x_\beta+\bar\partial_\beta)(Q_0-2)x.\psi.x-4(3x_\alpha+\bar\partial_\alpha)(Q_0-2)x_\beta
x.\psi.x.\label{34-5}
\end{equation}
Therefore, an initial equation is achieved to demonstrate the
effect of $(Q_0-2)Q_0$ on $k_\alpha\beta$.
\begin{equation}
(Q_0-2)Q_0k_{\alpha\beta}=-4x_\alpha(3x_\beta+\bar\partial_\beta)(Q_0-2)x.\psi.x-4(3x_\alpha+\bar\partial_\alpha)(Q_0-2)\psi_\alpha.x
$$$$-(3x_\alpha+\bar\partial_\alpha)(Q_0-2)\bar\partial.k_\beta.\label{35-5}
\end{equation}
If (\ref{35-5}) is multiplied by $x_\beta$, we will have
\begin{equation}
4(Q_0-2)\bar\partial.k_\alpha=12(Q_0-2)\psi_\alpha.x-3x_\alpha(2\bar\partial.\bar\partial.k)-2\bar\partial_\alpha\bar\partial.\bar\partial.k$$$$
+(Q_0-2)\bar\partial.k_\alpha+x_\alpha(2\bar\partial.\bar\partial.k)+x_\alpha(Q_0-2)\bar\partial.\bar\partial.k.\label{36-5}
\end{equation}
Therefore
\begin{equation}
(Q_0-2)\psi_\alpha.x=\frac{1}{4}(Q_0-2)\bar\partial.k_\alpha+\frac{1}{3}x_\alpha\bar\partial.\bar\partial.k-\frac{1}{12}x_\alpha(Q_0-2)\bar\partial.\bar\partial.k+\frac{1}{6}
\bar\partial_\alpha\bar\partial.\bar\partial.k. \label{37-5}
\end{equation}
Substitution in (\ref{35-5}) yields
\begin{equation}
(Q_0-2)Q_0k_{\alpha\beta}=-Q_0Sx_\beta\bar\partial\cdot
k_\alpha-Q_0S\bar\partial_\alpha\bar\partial\cdot
k_\beta+2Sx_\alpha\bar\partial\cdot
k_\beta$$$$+2S\bar\partial_\alpha\bar\partial\cdot
k_\beta-4x_\alpha x_\beta \bar\partial\cdot\bar\partial\cdot
k-\frac{1}{3}S\bar\partial_\beta\bar\partial_\alpha\bar\partial\cdot\bar\partial\cdot
k$$$$-\frac{5}{3}Sx_\alpha\bar\partial_\beta\bar\partial\cdot\bar\partial\cdot
k-2\theta_{\alpha\beta}\bar\partial\cdot\bar\partial\cdot
k+\frac{1}{3}\theta_{\alpha\beta}Q_0\bar\partial\cdot\bar\partial\cdot
k.
\label{38-5}
\end{equation}
And since
\begin{equation}
-\frac{1}{3}S\bar\partial_\beta\bar\partial_\alpha =
-\frac{2}{3}\bar\partial_\beta\bar\partial_\alpha +
\frac{1}{3}x_\alpha\bar\partial_\beta
-\frac{1}{3}x_\beta\bar\partial_\alpha, \label{39-5}
\end{equation}
we obtain the following relation
\begin{equation}
(Q_2+4)[(Q_2+6)k_{\alpha\beta}+D_2\bar\partial\cdot
k_\alpha]+\frac{1}{3}D_2D_1\bar\partial\cdot\bar\partial\cdot
k-\frac{1}{3}\theta_{\alpha\beta}(Q_0+6)\bar\partial\cdot\bar\partial\cdot
k=0.
\label{40-5}
\end{equation}
Equation (\ref{40-5}) is the conformally invariant equation of
conformal degree 2, for the second ranked tensor $k_{\alpha\beta}$
\cite{40}.

\addcontentsline{toc}{chapter}{Conclusion}
\include{Conclusions}
\addcontentsline{toc}{chapter}{Appendix A}

\appendix
\include{appendix1}

\addcontentsline{toc}{chapter}{Appendix B}
\include{appendix2}

\addcontentsline{toc}{chapter}{Bibliography}
\bibliographystyle{plain}

\end{document}

%% file: acknowledgements.tex
{\it{\large{I am grateful to my supervisor Dr. Mohammad Reza
Tanhayi for the helps, supports and scientific training, during
this work and thereafter.}}}

%% file: Introduction.tex
{\textbf{\Large{Introduction}}}\\

Our intention in this thesis, pursuing Dirac's work, is to obtain
a conformally invariant equation, to describe graviton. Evidently,
the Einstein equation is a peculiar one, which in addition to its
simplicity in mathematical form, could well describe important
astrophysical observations (at least in their classical limits).
Nevertheless, this equations has been confronted some ambiguities;
for example when an acceptable explanation of Dark Matter or Dark
Energy is demanded. Some physicists believe that by substituting
the Einstein equation by an alternative one, or finding a
generalized theory of gravitation, this problem can be resolved.
Recently some interesting theories have been proposed, like $f(R)$
gravity, brane-world gravity, Lovelock gravity and etc., which all
of them can be categorized in modified gravitational models.\\

Indeed our attempts are supposed to result in Maxwell-like
equations for gravitation. As we know, the electromagnetic theory
is describing a vector boson, namely the photon. This spin-$1$
boson which travels with the speed of light, contains no mass. One
important trait of electromagnetic equations is the conformal
invariance, which is in harmony with the massless-ness of the
described boson, namely the photon. The Einstein equation are also
describing a massless spin-$2$ boson, namely the graviton. Hence,
the gravitational waves should travel with the speed of light.
However, Einstein equations which are supposed to describe the
graviton, lake in conformal invariance.

It is worth to note that in this thesis, we consider a tensorial
field to describe graviton. What is done in this work is to review
the paper in \cite{40}. Our method is the one which has been
demonstrated by Dirac, called the Dirac's six cone.

Historically, Dirac tried to find a conformally invariant equation
in Minkowski space. He considered a flat six-dimensional space due
to the six dimensions of the conformal group $SO(2,4)$.
Afterwards, he derived the field equations in the so-called space,
and projected them on a five-dimensional hyper-surface. Finally he
projected them again on the flat four-dimensional Minkowski space.

In this thesis however, after projecting the six-dimensional field
equations on the five-dimensional hyper-surface, applying some
transverse projector operators, called the induced metric, we will
project the five-dimensional equations on a four-dimensional
curved de Sitter space. The importance of de Sitter space is
because of its ability to being in consistent with the recent
astrophysical data (like accelerated expansion of the universe),
since it possesses a constant (the cosmological constant),
correlated to the vacuum energy theory. Therefore the final curved
background in this thesis, would be the de Sitter spacetime.

Moreover, the transverse projector operators will be introduced as
follows:
$$\theta_{\alpha\beta}=\eta_{\alpha\beta}+H^2x_\alpha x_\beta.$$
Using this operator, we project all the operators like
differentiation operators or Casimir operators, from a flat
five-dimensional space, on a four-dimensional curved space.
Therefore we shall project the equations, derived by the Dirac's
cone, on the de Sitter space. Although the final equations have
the conformal invariance, but they are still incapable to be
transformed by the irreducible representations of the conformal
group and therefore do not have physical interpretations.

%% file: Conclusions.tex
{\textbf{\Large{Conclusion}}}\\

The obtained equation in (\ref{40-5}), is a conformally invariant
equation for graviton. As we know, graviton is an spin-$2$
elementary particle, described by Einstein field equations. Since
this particle is supposed to be massless, it must be described by
a conformally invariant equation, which is the property that
Einstein equation lacks. The attempts in this thesis, were to
obtain a conformally invariant equation to describe graviton, or a
tensorial field of rank 2. These attempts led to equation
(\ref{40-5}).

Despite its conformal characteristics, this equation however does
not have physical descriptions, since it is not transformed by
irreducible representations of conformal group $SO(2,4)$. Such
problem, occurs for Weyl gravity, which is regarded as a
gravitational theory of higher order with respect to Ricci scalar.

Therefore it is considerable to relate a tensor of higher rank to
graviton, and try to find the conformal equation using this new
tensorial field. Pursuing this assumption, led to obtaining an
equation, which beside its conformal invariance, is capable to be
transformed be the conformal group \cite{41}. Theoretical
physicists are still trying to find the corresponding conformally
invariant Lagrangian.

%% file: appendix1.tex
{\textbf{\Large{Appendix A}}}\\

{\textbf{\large{Proof of important relations in chapter 4}}}\\

Proof of relation (\ref{12-4}):
$$\bar\partial_\alpha H^{-2}=[\partial_\alpha+H^2x_\alpha(x.\partial)]H^{-2}$$
$$\partial_\alpha H^{-2}+H^2
x_\alpha(x.\partial)H^{-2}-2x_\alpha+H^2x_\alpha(x_\beta\partial^\beta)H^{-2}$$
$$=-2x_\alpha-2H^{2}x_\alpha x_\beta x^\beta=-2x_\alpha-2x_\alpha H^2(-H^{-2}) $$
$$=-2x_\alpha+2x_\alpha=0.$$

Proof of relation (\ref{13-4}):
$$\partial_\alpha H^{-2}=\partial_\alpha(-x^\beta x_\beta)=-2x_\alpha.$$

Proof of relation (\ref{14-4}):
$$\bar\partial_\alpha x^\beta=[\partial_\alpha+H^2 x_\alpha(x.\partial)]x^\beta$$
$$=\eta^\beta_\alpha+H^2x_\alpha x^\gamma\partial_\gamma x^\beta$$
$$=\eta^\beta_\alpha+H^2x_\alpha x^\beta\equiv\theta^\beta_\alpha,\,\,\,\,\,\,\,\,\,\,\,\,\,\,\partial_\gamma x^\beta=\eta^\beta_\alpha.$$

Proof of relation (4.17):
$$x_\beta\bar\partial_\beta-x_\beta\bar\partial_\alpha=x_\alpha[\partial_\beta+H^2 x_\beta(x.\partial)]-x_\beta[\partial_\alpha+H^2(x.\partial)]$$
$$=x_\alpha\partial_\beta-x_\beta\partial_\alpha+H^2 x_\alpha x_\beta(x.\partial)-H^2x_\beta x_\alpha(x.\partial)$$
$$=x_\alpha\partial_\beta-x_\beta\partial_\alpha\,\,\,\,\,\,\,\,\,\,\,\Rightarrow\,\,\,\,\,\,\,\,\,\,x^\alpha\bar\partial^\beta-x^\beta\bar\partial^\alpha=x^\alpha\partial^\beta-x^\beta\partial^\alpha.$$

Proof of relation (\ref{19-4}):
$$-\frac{1}{2}M_{\alpha\beta}M^{\alpha\beta}=-\frac{1}{2}[x_\alpha\bar\partial_\beta-x_\beta\bar\partial_\alpha]
[x^\alpha\bar\partial^\beta-x^\beta \bar\partial^\alpha]$$
$$-\frac{1}{2}[x_\alpha\bar\partial_\beta x^\alpha\bar\partial^\beta-x_\alpha\bar\partial_\beta x^\beta\bar\partial^\alpha
-x_\beta\bar\partial_\alpha
x^\alpha\bar\partial^\beta+x_\beta\bar\partial_\alpha
x^\beta\bar\partial^\alpha]$$
$$=-\frac{1}{2}[x_\alpha\theta^\alpha_\beta\bar\partial^\beta-x^2\bar\partial_\beta\bar\partial^\beta-x_\alpha(\bar\partial.x)\bar\partial^\alpha
-x_\alpha x^\beta\bar\partial_\beta\bar\partial^\alpha$$
$$-x_\beta(\bar\partial.x)\bar\partial^\beta+x_\beta x^\alpha\bar\partial_\alpha\bar\partial^\beta+x_\beta\theta^\beta_\alpha\bar\partial^\alpha-x_\beta x^\beta\bar\partial_\alpha\bar\partial^\alpha$$
$$=H^{-2}\bar\partial_\alpha\bar\partial^\alpha=H^{-2}\bar\partial^2.$$

Proof of relation (\ref{20-4}):
$$\theta_{\alpha\beta}\theta^{\alpha\gamma}=(\eta_{\alpha\beta}+H^2 x_\alpha x_\beta)(H^2 x^\alpha x^\gamma)$$
$$=\eta_{\alpha\beta}\eta^{\alpha\gamma}+H^2(x_\alpha x^\gamma+x^\gamma x_\beta)+H^2 x_\alpha x^\beta x_\beta x^\gamma$$
$$=\eta_{\alpha\beta}\eta^{\alpha\gamma}+2H^2x_\beta x^\gamma-H^2 x_\beta x^\gamma=\eta^\alpha_\beta+H^2 x_\beta x^\gamma=\theta^\alpha_\beta.$$

Proof of relation (\ref{21-4}):
$$\bar\partial_\beta\theta_{\alpha\sigma}=\bar\partial_\beta(\eta_{\alpha\sigma}+H^2x_\alpha x_\sigma)=\bar\partial_\beta(H^2 x_\alpha x_\sigma)
=H^2\theta_{\beta\alpha} x_\sigma+H^2 x_\sigma
\theta_{\beta\sigma}.$$

Proof of relation (\ref{24-4}):
$$\bar\partial_\alpha\bar\partial_\beta=\bar\partial_\alpha(\theta_{\beta\gamma}\partial_\gamma)=(\bar\partial_\alpha\theta_{\beta\gamma})\partial_\gamma
+\theta_{\beta\gamma}\bar\partial_\alpha\partial_\gamma$$
$$=\theta_{\beta\gamma}\theta_{\alpha\lambda}\partial_\gamma\partial_\lambda+H^2\theta_{\alpha\gamma} x_\beta\delta^\alpha_\beta\partial_\gamma+H^2\theta_{\beta\gamma} x_\gamma\delta^\alpha_\gamma\partial_\gamma$$
$$=\theta_{\beta\gamma}\theta_{\alpha\lambda}\partial_\gamma\partial_\lambda+(H^2\theta_{\alpha\gamma}x_\beta+
H^2\theta_{\alpha\beta}x_\gamma)\partial_\gamma$$
$$=\theta_{\beta\gamma}\theta_{\alpha\lambda}\partial_\gamma\partial_\lambda+H^2x_\beta \bar\partial_\alpha
+H^2\theta_{\alpha\beta}(x.\partial)$$
$$=\partial_\beta\bar\partial_\alpha-H^2\theta_{\beta\gamma}-H^2\theta_{\beta\gamma}[\delta_{\gamma\alpha} x_\lambda
+\delta_{\gamma\lambda}x_\alpha]\partial_\lambda+H^2[x_\beta\bar\partial_\alpha+\theta_{\alpha\beta}(x.\partial)]$$
$$\Rightarrow\,\,\,\,\,\,\,\,\,\,\bar\partial_\alpha\bar\partial_\beta-\bar\partial_\beta\bar\partial_\alpha=-H^2 x_\alpha\bar\partial_\beta+H^2x_\beta\bar\partial_\alpha$$
$$\Rightarrow\,\,\,\,\,\,\,\,\,\,[\bar\partial_\alpha,\bar\partial_\beta]=H^2(x_\beta\bar\partial_\alpha-x_\alpha\bar\partial_\beta).$$

Proof of relation (\ref{25-4}):
$$Q_0 x_\alpha f=-H^{-2}\bar\partial^\gamma\bar\partial_\gamma x_\alpha f=-H^{-2}\bar\partial^\gamma[\theta_{\gamma\alpha}f+x_\alpha\bar\partial_\gamma f]$$
$$=-H^{-2}[(\bar\partial^\gamma\theta_{\gamma\alpha})f+\theta_{\gamma\alpha}\bar\partial^\gamma f+(\bar\partial^\gamma x_\alpha)\bar\partial_\gamma f
x_\alpha\bar\partial^\alpha\bar\partial_\gamma f]$$
$$\Rightarrow\,\,\,\,\,\,\,\,\,\,Q_0 x_\alpha=-4x_\alpha-2H^{-2}\bar\partial_\alpha+x_\alpha Q_0$$
$$\Rightarrow\,\,\,\,\,\,\,\,\,\,[Q_0,x_\alpha]=-4x_\alpha-2H^{-2}\bar\partial_\alpha.$$

Proof of relation (\ref{26-4}):
$$Q_0\bar\partial_\alpha f=-H^{-2}\bar\partial^\beta\bar\partial_\beta(\bar\partial_\alpha f)=-H^{-2}\bar\partial^\beta\bar\partial_\beta\bar\partial_\alpha f$$
$$=-H^{-2}\bar\partial^\beta[\bar\partial_\alpha\bar\partial_\beta-H^2 x_\beta\bar\partial_\alpha+H^2 x_\alpha\bar\partial_\beta]$$
$$=-H^{-2}[\bar\partial_\alpha\bar\partial^\beta-H^2x^\beta\bar\partial_\alpha+H^2x_\alpha\bar\partial^\beta]+(\bar\partial^\beta x_\beta)\bar\partial_\alpha$$
$$+x_\beta\bar\partial^\beta\bar\partial_\alpha-[({\bar\partial^\beta x_\alpha})\bar\partial_\beta    +x_\alpha\bar\partial^\beta\bar\partial_\beta]$$
$$=\bar\partial_\alpha Q_0+(\bar\partial_\alpha x_\beta-\theta^\beta_\alpha)\bar\partial_\beta+2H^2(Q_0x_\alpha+4x_\alpha+2H^{-2}\bar\partial_\alpha)+3\bar\partial_\alpha$$
$$\Rightarrow\,\,\,\,\,\,\,\,\,\,Q_0\bar\partial_\alpha=\bar\partial_\alpha Q_0+6\bar\partial_\alpha+2H^2(Q_0+4)x_\alpha$$
$$\Rightarrow\,\,\,\,\,\,\,\,\,\,[Q_0,\bar\partial_\alpha]=6\bar\partial_\alpha+2H^2(Q_0+4)x_\alpha.$$

Proof of relation (\ref{30-4}):
$$
D_1Q_0\phi=H^{-2}\bar\partial_\alpha
Q_0\phi=H^{-2}Q_0\bar\partial_\alpha\phi-6H^{-2}\bar\partial_\alpha\phi-2(Q_0+4)x_\alpha\phi$$
$$=H^{-2}Q_0\bar\partial_\alpha\phi-6H^{-2}\bar\partial_\alpha\phi-8x_\alpha\phi-2x_\alpha Q_0\phi+8x_\alpha\phi+4H^{-2}\bar\partial_\alpha\phi$$
$$\Rightarrow\,\,\,\,\,\,\,\,\,\,D_1Q_0\phi=H^{-2}Q_0\bar\partial_\alpha\phi-2H^{-2}\bar\partial_\alpha\phi-8x_\alpha\phi-2x_\alpha
Q_0\phi. $$
 On the other hand
$$
Q_1D_{1\alpha}\phi=(Q_0-2)D_{1\alpha}\phi-2x_\alpha\bar\partial_\beta
D_{1\beta}\phi$$$$=Q_0D_1\phi-2D_1\phi-2x_\alpha Q_0\phi.
$$
Therefore
$$Q_1 D_1\phi=D_1Q_0\phi.$$

Proof of relation (\ref{36-4}):
$$Q_0x.z\phi=Q_0x^\alpha z_\alpha\phi=z_\alpha Q_0 x^\alpha\phi=z_\alpha[x_\alpha Q_0-4x_\alpha-2H^{-2}\bar\partial_\alpha]\phi$$
$$=(x.z)(Q_0-4)\phi-2H^{-2}z.\bar\partial\phi.$$

Proof of relation (\ref{53-4})
$$x.k=x^\alpha k_\alpha=x_5(x^\alpha\psi_\alpha+x^\alpha x_\alpha x.\psi)$$
$$x_5(x.\psi-x.\psi)=0.$$

%% file: appendix2.tex
{\textbf{\Large{Appendix B}}}\\

{\textbf{\large{Detailed mathematical calculations of chapter
5}}}\\

For $p = 2$ The Conformal system due to Dirac's six cone is
$$(Q_0-2)Q_0\psi_{AB}=0,$$$$\hat{N}_5\psi_{AB}=0.$$
Or
$$(Q_0-2)Q_0\psi'=0,$$
$$(Q_0-2)Q_0\psi_{\alpha\beta}=0,$$
$$(Q_0-2)Q_0\psi_{55}=0.$$
In general
$$
(Q_2+6)k_{\alpha\beta}=Q_0
k_{\alpha\beta}+2Sx_\alpha\bar\partial.\
k_\beta+2\eta_{\alpha\beta}k'-2S\bar\partial_\alpha x. k_\beta,$$
where
$$k_{\alpha\beta}=\psi_{\alpha\beta}+Sx_\beta\psi_\alpha.
x+x_\alpha x_\beta x.\psi. x.$$
The transversality condition yields
$$u^5\psi_{AB}=0
\,\,\,\,\,\,\,\,\,\,\Rightarrow\,\,\,\,\,\,\,\,\,\,
x^5(\psi_{5B}+x.\psi_B)=0$$
$$\Rightarrow\,\,\,\,\,\,\,\,\,\,(Q_0-2)Q_0\psi_{5B}+(Q_0-2)Q_0x.\psi_B=0$$
$$\Rightarrow\,\,\,\,\,\,\,\,\,\,(Q_0-2)Q_0x.\psi_B=0.$$
And the tracelessness implies
yields$$k'=0\Rightarrow\psi_{\alpha\alpha}+x.\psi. x=0.$$
Therefore
$$(Q_0-2)Q_0x.\psi_\beta=0\,\,\,\,\,\,\,\,\,\,\,\,\,\,\,\,\,\,\,\,(\textmd{I})$$$$(Q_0-2)Q_0x.\psi.
x=0   \,\,\,\,\,\,\,\,\,\,\,\,\,\,\,\,\,\,\,\,(\textmd{II})$$We
can show that
$$x_\alpha(Q_0-2)Q_0=(Q_0-2)(Q_0x_\alpha+4x_\alpha+4\bar\partial_\alpha)=(Q_0-2)Q_0x_\alpha+4(3x_\alpha+\bar\partial_\alpha)(Q_0-2).$$Multiply $x_\beta$
by
(I):$$(Q_0-2)(Q_0x_\beta+4x_\beta+4\bar\partial_\beta)x.\psi_\beta=0$$$$\Rightarrow\,\,\,\,\,\,\,\,\,\,(Q_0-2)Q_0x.\psi.
x+4(Q_0-2)x.\psi. x+4(Q_0-2)\bar\partial.\psi. x=0.$$From (II) we
have$$(Q_0-2)x.\psi. x=-(Q_0-2)\bar\partial.\psi.
x$$$$\Rightarrow\,\,\,\,\,\,\,\,\,\,(Q_0-2)(x.\psi.
x+\bar\partial.\psi.
x)=0.\,\,\,\,\,\,\,\,\,\,\,\,\,\,\,\,\,\,\,\,(\textmd{III})$$
 The divergence of $\psi_{\alpha\beta}$
is$$\nabla_a \psi^{ab}=0,$$ therefore$$\partial.\psi_B=-x.\partial
x.\psi_B,$$and$$\partial.\psi_5=-x.\partial x.\psi_5.$$Also the
transversality condition $u_a\psi^{ab}=0$ results in
$$x^5(\psi_{5b}+x.\psi_b)=0,$$
therefore$$\partial.\psi. x +x.\partial x.\psi. x=0.$$The
divergence of $k_{\alpha\beta}$ is
$$\bar\partial. k_\alpha=\bar\partial.\psi_\alpha+4\psi_\alpha. x+\theta_{\alpha\beta}\psi_\beta.
x+x_\alpha\bar\partial.\psi. x+\theta_{\alpha\beta}x_\beta x.\psi.
x+4x_\alpha x.\psi.
x$$$$\Rightarrow\,\,\,\,\,\,\,\,\,\,\bar\partial.
k_\alpha=4(\psi_\alpha. x+x_\alpha x.\psi.
x)+(\bar\partial.\psi_\alpha+x.\psi_\alpha+x_\alpha\bar\partial.\psi.
x +x_\alpha x.\psi. x ).$$ Since our conformal degree degree is 1,
we
have:
$$\bar\partial.\psi_\beta=-x.\psi_\beta,$$and$$\bar\partial.\psi.
x=-x.\psi. x.$$ Therefore$$\bar\partial. k_\alpha=4(\psi_\alpha.
x+x_\alpha x.\psi.
x).\,\,\,\,\,\,\,\,\,\,\,\,\,\,\,\,\,\,\,\,(\textmd{IV})$$Let us
combine (III) and (IV). This yields
have$$(Q_0-2)(\frac{1}{12}\bar\partial.\bar\partial. k-x.\psi.
x)=0.\,\,\,\,\,\,\,\,\,\,\,\,\,\,\,\,\,\,\,\,(\textmd{V})$$ Now
multiply both sides of equation (I) by $\bar\partial_\beta$. We
obtain$$(Q_0-2)Q_0\bar\partial.\psi.\
x-4x_\beta(Q_0-2)Q_0x.\psi_\beta=0.$$ Using (I) we
have$$(Q_0-2)Q_0\bar\partial.\psi. x=0.$$The effect of
$(Q_0-2)Q_0$ on our tensor field, will
be
$$(Q_0-2)Q_0k_{\alpha\beta}=(Q_0-2)Q_0Sx_\alpha\psi_\beta.
x+(Q_0-2)Q_0x_\alpha x_\beta x.\psi.
x.\,\,\,\,\,\,\,\,\,\,\,\,\,\,\,\,\,\,\,\,(\star)$$ We
have$$(Q_0-2)Q_0x_\alpha\psi_\beta.
x=x_\alpha(Q_0-2)Q_0\psi_\beta.
x-4(3x_\alpha+\bar\partial_\alpha)(Q_0-2)x_\beta x.\psi.
x,$$and$$(Q_0-2)Q_0x_\alpha x_\beta x.\psi. x=x_\alpha
x_\beta(Q_0-2)Q_0x.\psi.
x-4x_\alpha(3x_\beta+\bar\partial_\beta)(Q_0-2)x.\psi.
x$$$$-4(3x_\alpha+\bar\partial_\alpha)(Q_0-2)x_\beta x.\psi.
x.$$Therefore using equation (IV), equation $(\star)$ by
substitution
yields$$(Q_0-2)Q_0k_{\alpha\beta}=-4(3x_\alpha+\bar\partial_\alpha)(Q_0-2)\psi_\alpha.
x-4x_\alpha(3x_\beta+\bar\partial_\beta)(Q_0-2)x.\psi.\
x$$$$-(3x_\alpha+\bar\partial_\alpha)(Q_0-2)\bar\partial.
k_\beta.$$Finally using equation (V) we
get$$(Q_0-2)k_{\alpha\beta}=-4(3x_\beta+\bar\partial_\beta)(Q_0-2)\psi_\alpha.
x-(3x_\alpha+\bar\partial_\alpha)(Q_0-2)\bar\partial.
k_\beta$$$$-\frac{1}{3}x_\alpha(3x_\beta+\bar\partial_\beta)(Q_0-2)\bar\partial.\bar\partial.
k.\,\,\,\,\,\,\,\,\,\,\,\,\,\,\,\,\,\,\,\,(\star\star)$$Now let us
multiply both sides of above equation by
$x_\beta$.$$4(Q_0-2)\bar\partial. k_\alpha=12(Q_0-2)\psi_\alpha.
x-3x_\alpha x_\beta(Q_0-2)\bar\partial.
k_\beta-\bar\partial_\alpha x_\beta(Q_0-2)\bar\partial.
k_\beta$$$$-\theta_{\alpha\beta}(Q_0-2)\bar\partial.
k_\beta+x_\alpha(Q_0-2)\bar\partial.\bar\partial.
k$$$$\Rightarrow\,\,\,\,\,\,\,\,\,\,4(Q_0-2)\bar\partial.
k_\alpha=12(Q_0-2)\psi_\alpha.
x-3x_\alpha(2\bar\partial.\bar\partial.
k)-2\bar\partial_\alpha\bar\partial.\bar\partial.
k$$$$+(Q_0-2)\bar\partial.
k_\alpha+x_\alpha(2\bar\partial.\bar\partial.
k)+x_\alpha(Q_0-2)\bar\partial.\bar\partial.
k$$$$\Rightarrow\,\,\,\,\,\,\,\,\,\,(Q_0-2)\psi_\alpha.
x=\frac{1}{4}(Q_0-2)\bar\partial.
k_\alpha+\frac{1}{3}x_\alpha\bar\partial.\bar\partial.
k-\frac{1}{12}x_\alpha(Q_0-2)\bar\partial.\bar\partial.
k+\frac{1}{6}\bar\partial_\alpha\bar\partial.\bar\partial. k.$$
Substitution in $(\star\star)$
yields$$(Q_0-2)Q_0k_{\alpha\beta}=-4(3x_\beta+\bar\partial_\beta)\{\frac{1}{4}(Q_0-2)\bar\partial.
k_\alpha+\frac{1}{3}x_\alpha\bar\partial.\bar\partial.
k-\frac{1}{12}x_\alpha(Q_0-2)\bar\partial.\bar\partial.
k+\frac{1}{6}\bar\partial_\alpha\bar\partial.\bar\partial.
k\}$$$$-(3x_\alpha+\bar\partial_\alpha)(Q_0-2)\bar\partial.
k_\beta-\frac{1}{3}x_\alpha(3x_\beta+\bar\partial_\beta)(Q_0-2)\bar\partial.\bar\partial.
k$$$$\Rightarrow\,\,\,\,\,\,\,\,\,\,(Q_0-2)Q_0k_{\alpha\beta}=-(3x_\beta+\bar\partial_\beta)(Q_0-2)\bar\partial.
k_\alpha-\frac{4}{3}(3x_\beta+\bar\partial_\beta)x_\alpha\bar\partial.\bar\partial.
k$$$$+\frac{1}{3}(3x_\beta+\bar\partial_\beta)x_\alpha(Q_0-2)\bar\partial.\bar\partial.
k-\frac{2}{3}(3x_\beta+\bar\partial_\beta)\bar\partial_\alpha\bar\partial.\bar\partial.
k$$$$-(3x_\alpha+\bar\partial_\alpha)(Q_0-2)\bar\partial.
k_\beta-\frac{1}{3}x_\alpha(3x_\beta+\bar\partial_\beta)(Q_0-2)\bar\partial.\bar\partial.
k$$$$\Rightarrow\,\,\,\,\,\,\,\,\,\,(Q_0-2)Q_0k_{\alpha\beta}=-3x_\beta(Q_0-2)\bar\partial.
k_\alpha-\bar\partial_\beta(Q_0-2)\bar\partial. k_\alpha-4x_\beta
x_\alpha\bar\partial.\bar\partial.
k$$$$-\frac{4}{3}\bar\partial_\beta
x_\alpha\bar\partial.\bar\partial. k+x_\beta
x_\alpha(Q_0-2)\bar\partial.\bar\partial.
k+\frac{1}{3}\bar\partial_\beta
x_\alpha(Q_0-2)\bar\partial.\bar\partial.
k-2x_\beta\bar\partial_\alpha\bar\partial.\bar\partial.
k$$$$-\frac{2}{3}\bar\partial_\beta\bar\partial_\alpha\bar\partial.\bar\partial.
k-3x_\alpha(Q_0-2)\bar\partial.
k_\beta-\bar\partial_\alpha(Q_0-2)\bar\partial.
k_\beta$$$$-x_\alpha x_\beta(Q_0-2)\bar\partial.\bar\partial. k
-\frac{1}{3}x_\alpha
\bar\partial_\beta(Q_0-2)\bar\partial.\bar\partial. k.$$Extending
the operators on the vectors and partial differentials, leads
to$$(Q_0-2)Q_0k_{\alpha\beta}=-3Q_0x_\beta\bar\partial.
k_\alpha-6x_\beta\bar\partial.
k_\alpha-6\bar\partial_\beta\bar\partial.
k_\alpha-Q_0\bar\partial_\beta\bar\partial.
k_\alpha+8\bar\partial_\beta\bar\partial.
k_\alpha$$$$+2Q_0x_\beta\bar\partial.
k_\alpha+8x_\beta\bar\partial. k_\alpha-4x_\alpha
x_\beta\bar\partial.\bar\partial.
k-\frac{4}{3}\theta_{\alpha\beta}\bar\partial.\bar\partial.
k$$$$-\frac{4}{3}x_\alpha\bar\partial_\beta\bar\partial.\bar\partial.
k+\frac{1}{3}\theta_{\alpha\beta}(Q_0-2)\bar\partial.\bar\partial.
k+\frac{1}{3}x_\alpha\bar\partial_\beta(Q_0-2)\bar\partial.\bar\partial.
k$$$$-2x_\beta\bar\partial_\alpha\bar\partial.\bar\partial.
k-\frac{2}{3}\bar\partial_\beta\bar\partial_\alpha\bar\partial.\bar\partial.
k-3Q_0x_\alpha\bar\partial. k_\beta-6x_\alpha\bar\partial.
k_\beta$$$$-6\bar\partial_\alpha\bar\partial.
k_\beta-Q_0\bar\partial_\alpha\bar\partial.
k_\beta+8\bar\partial_\alpha\bar\partial.
k_\beta+2Q_0x_\alpha\bar\partial.
k_\beta$$$$+8x_\alpha\bar\partial.
k_\beta-\frac{1}{3}x_\alpha\bar\partial_\beta
Q_0\bar\partial.\bar\partial.
k+\frac{2}{3}x_\alpha\bar\partial_\beta\bar\partial.\bar\partial.
k.$$After some simplifications, we will have
$$(Q_0-2)Q_0
k_{\alpha\beta}=-Q_0x_\beta\bar\partial.
k_\alpha-Q_0x_\alpha\bar\partial.
k_\beta-Q_0\bar\partial_\beta\bar\partial.
k_\alpha-Q_0\bar\partial_\alpha\bar\partial.
k_\beta$$$$+2x_\beta\bar\partial. k_\alpha+2x_\alpha\bar\partial.
k_\alpha+2\bar\partial_\beta\bar\partial.
k_\alpha+2\bar\partial_\alpha\bar\partial. k_\alpha$$$$-4x_\alpha
x_\beta\bar\partial.\bar\partial.
k-\frac{4}{3}x_\alpha\bar\partial_\beta\bar\partial.\bar\partial.
k-2x_\beta\bar\partial_\alpha\bar\partial.\bar\partial.
k$$$$-2\theta_{\alpha\beta}\bar\partial.\bar\partial.
k+\frac{1}{3}\theta_{\alpha\beta}Q_0\bar\partial.\bar\partial. k
-\frac{2}{3}\bar\partial_\beta\bar\partial_\alpha\bar\partial.\bar\partial.
k.$$Finally, the conformal equation for tensor fields, with $p =
2$, is derived,
i.e.$$(Q_0-2)Q_0k_{\alpha\beta}=-Q_0Sx_\beta\bar\partial.
k_\alpha-Q_0S\bar\partial_\alpha\bar\partial.
k_\beta+2Sx_\alpha\bar\partial.
k_\beta$$$$+2S\bar\partial_\alpha\bar\partial. k_\beta-4x_\alpha
x_\beta \bar\partial.\bar\partial.
k-\frac{1}{3}S\bar\partial_\beta\bar\partial_\alpha\bar\partial.\bar\partial.
k$$$$-\frac{5}{3}Sx_\alpha\bar\partial_\beta\bar\partial.\bar\partial.
k-2\theta_{\alpha\beta}\bar\partial.\bar\partial.
k+\frac{1}{3}\theta_{\alpha\beta}Q_0\bar\partial.\bar\partial.
k.\,\,\,\,\,\,\,\,\,\,\,\,\,\,\,\,\,\,\,\,(\star\star\star)$$
Note
that$$-\frac{1}{3}S\bar\partial_\beta\bar\partial_\alpha=\frac{1}{3}(\bar\partial_\beta\bar\partial_\alpha+\bar\partial_\alpha\bar\partial_\beta)
=-\frac{1}{3}(2\bar\partial_\beta\bar\partial_\alpha-x_\alpha\bar\partial_\beta+x_\beta\bar\partial_\alpha)$$
$$\Rightarrow\,\,\,\,\,\,\,\,\,\,-\frac{1}{3}S\bar\partial_\beta\bar\partial_\alpha = -\frac{2}{3}\bar\partial_\beta\bar\partial_\alpha + \frac{1}{3}x_\alpha\bar\partial_\beta
-\frac{1}{3}x_\beta\bar\partial_\alpha.$$Another representation of
$(\star\star\star)$
is$$(Q_2+4)[(Q_2+6)k_{\alpha\beta}+D_2\bar\partial.
k_\alpha]+\frac{1}{3}D_2D_1\bar\partial.\bar\partial.
k-\frac{1}{3}\theta_{\alpha\beta}(Q_0+6)\bar\partial.\bar\partial.
k=0.$$
\\\\

Now let us prove the recursion relation between (\ref{20-5}) and
(\ref{22-5}). The transverse projections of all the parts of Barut
and Xu conformal equation, are as below
$$\Box h_{\mu\nu}= \textmd{TrPr} \bar\partial_\alpha \bar\partial_\alpha
k_{\mu\nu}= \bar\partial_\alpha (\bar\partial_\alpha k_{\mu\nu} -
x_\mu k_{\alpha\nu}- x_\nu k_{\mu\alpha})-
x_\alpha(\bar\partial_\alpha k_{\mu\nu}- x_\mu k_{\alpha\nu}-
x_\nu k_{\mu\alpha})$$
$$- x_\mu(\bar\partial_\alpha k_{\alpha\nu}- x_\alpha
k_{\alpha\nu}-x_\nu k_{\alpha\alpha})-x_\nu(\bar\partial_\alpha
k_{\mu\alpha}-x_\mu k _{\alpha\alpha}- x_\alpha k _{\mu\alpha})$$
$$ = \bar\partial_\alpha \bar\partial_\alpha k_{\mu\nu} -
 \theta_{\alpha\mu}k_{\alpha\nu}-\theta_{\alpha\nu}k_{\mu\alpha}-x_\mu
 \bar\partial_\alpha k_{\alpha\nu}-x_\nu \bar\partial_\alpha
 k_{\mu\alpha}$$
$$ =(Q_0 - 2)k_{\mu\nu}- 2 S x_\mu \bar\partial\ k_\nu.\,\,\,\,\,\,\,\,\,\,\,\,\,\,\,\,\,\,\,\,(\textmd{VI})$$

 $$\nabla_\mu\nabla\ h_\nu = \textmd{TrPr}
 \bar\partial_\mu\bar\partial_\lambda k_{\lambda\nu}=
 \bar\partial_\mu(\bar\partial_\lambda k_{\lambda\nu}-x_\lambda
 k_{\lambda\nu}-x_\nu
 k_{\lambda\lambda})-x_\lambda(\bar\partial_\mu
 k_{\lambda\nu}-x_\lambda k_{\mu\nu}- x_\nu k_{\lambda\mu})$$
 $$-x_\lambda(\bar\partial\lambda k_{\mu\nu}- x_\mu
 k_{\lambda\nu}-x_\nu k_{\mu\lambda})
 -x_\nu(\bar\partial_\lambda k _{\mu\lambda}-x_\lambda
 k_{\lambda\mu}-x_\mu k _{\lambda\lambda})$$
$$=\bar\partial_\mu\bar\partial_\lambda k_{\lambda\nu}+ x_\lambda
x_\lambda k_{\mu\nu}- x_\lambda \bar\partial_\mu k _{\lambda\nu} -
x_\nu\bar\partial_\lambda k _{\mu\lambda}$$
$$-\bar\partial_\mu\bar\partial\ k_\nu - x_\nu\bar\partial k _\nu.
\,\,\,\,\,\,\,\,\,\,\,\,\,\,\,\,\,\,\,\,(\textmd{VII})$$

$$\nabla\nabla h = \textmd{TrPr}\bar\partial_\mu \textmd{TrPr}\bar\partial_\nu
=\bar\partial_\mu(\bar\partial_\nu k_{\mu\nu}-x_\mu
 k_{\nu\nu}-x_\nu
 k_{\mu\nu})-x_\nu(\bar\partial_\mu
 k_{\mu\nu}-x_\mu k_{\mu\nu}- x_\nu k_{\mu\mu})$$
 $$-x_\mu(\bar\partial\nu k_{\mu\nu}- x_\mu
 k_{\nu\nu}-x_\nu k_{\mu\nu})
 -x_\nu(\bar\partial_\nu k _{\mu\mu}-x_\mu
 k_{\mu\nu}-x_\mu k _{\mu\nu}) $$
$$ =\bar\partial_\mu\bar\partial_\nu k_{\mu\nu} - \theta_{\mu\nu}
k_{\mu\nu}= \bar\partial\bar\partial k - k_{\mu\nu}.
 \,\,\,\,\,\,\,\,\,\,\,\,\,\,\,\,\,\,\,\,(\textmd{VIII})$$

 Note the following transformation:
 $$g^{\textmd{\small{ds}}}_{\mu\nu} \longmapsto \theta_{\mu\nu}\,\,\,\,\,\,\,\,\,\,\,\,\,\,\,\,\,\,\,\,(\textmd{IX})$$
 Therefore
 $$(\Box + 4)h_{\mu\nu}-\frac {2}{3}S\nabla_\mu\nabla
 h_\nu+\frac {1}{3}g^{ds}_{\mu\nu}\nabla\nabla h = 0.
$$

$$ (Q_0 - 2)k_{\mu\nu}-2Sx_\mu\bar\partial
k_\nu-\frac{2}{3}[S\bar\partial_\mu\bar\partial
k_\nu-Sx_\nu\bar\partial
k_\mu]+\frac{1}{3}\theta_{\mu\nu}[\bar\partial\bar\partial
k-k_{\mu\nu}]= 0 $$
$$ \Rightarrow\,\,\,\,\,\,\,\,\,\,(Q_0
-2)k_{\mu\nu}-\frac{2}{3}S(\bar\partial\mu+2x_\mu)\bar\partial
k_\nu+\frac{1}{3}\theta_{\mu\nu}\bar\partial\bar\partial k = 0.$$